\documentclass{aastex62}
\usepackage{tabularx}
\usepackage[hyphenbreaks]{breakurl}

\graphicspath{{./}{figures/}}

\received{}
\revised{}
\accepted{}
\submitjournal{ApJ}

\shorttitle{Compressive turbulence at sub-ion scales}
\shortauthors{Roberts et al.}

\begin{document}

\title{Sub-ion scale Compressive Turbulence in the Solar wind:  MMS spacecraft potential observations}

\correspondingauthor{Owen Wyn Roberts}
\email{owen.roberts@oeaw.ac.at}

\author[0000-0002-3913-1353]{Owen Wyn Roberts}
\affil{Space Research Institute, Austrian Academy of Sciences, Graz, Austria}

\author[0000-0002-2620-9211]{Rumi Nakamura}
\affiliation{Space Research Institute, Austrian Academy of Sciences, Graz, Austria}

\author[0000-0003-4820-1920]{Klaus Torkar}
\affiliation{Space Research Institute, Austrian Academy of Sciences, Graz, Austria}

\author{Yasuhito Narita}
\affiliation{Space Research Institute, Austrian Academy of Sciences, Graz, Austria}

\author[0000-0002-5414-8604]{Justin C. Holmes}
\affiliation{Space Research Institute, Austrian Academy of Sciences, Graz, Austria}

\author[0000-0001-7597-238X]{Zolt\'an V\"or\"os}
\affiliation{Space Research Institute, Austrian Academy of Sciences, Graz, Austria}
\affiliation{Geodetic and Geophysical Institute, Research Centre for Astronomy and Earth Sciences (RCAES), Sopron, Hungary}

\author[0000-0002-7552-2941]{Christoph Lhotka} 
\affiliation{Space Research Institute, Austrian Academy of Sciences, Graz, Austria}

\author[0000-0003-4475-6769]{C. Philippe Escoubet}
\affiliation{ESA, European Space Research and Technology Centre, Noordwijk, Netherlands}

\author[0000-0002-1046-746X]{Daniel B. Graham}
\affiliation{Swedish Institute of Space Physics, Uppsala, Sweden}


\author[0000-0003-1304-4769]{Daniel J. Gershman}
\affiliation{Goddard Space Flight Center, National Aeronautics and Space Administration, Greenbelt, MD, United States}

\author[0000-0001-5550-3113]{Yuri Khotyaintsev}
\affiliation{Swedish Institute of Space Physics, Uppsala, Sweden}

\author[0000-0001-5617-9765]{Per-Arne Lindqvist}
\affiliation{Department of Space and Plasma Physics, KTH Royal Institute of Technology, Stockholm, Sweden}



\begin{abstract}
Compressive plasma turbulence is investigated at sub-ion scales in the solar wind using both the Fast Plasma Investigation (FPI) instrument on the Magnetospheric MultiScale mission (MMS), as well as using calibrated spacecraft potential data from the Spin Plane Double Probe (SDP) instrument. The data from FPI allow the sub-ion scale region ($f_{sc}\gtrsim 1$ Hz) to be investigated before the instrumental noise becomes significant at a spacecraft frame frequency of $f_{sc}\approx 3$Hz. Whereas the calibrated spacecraft potential allows a measurement up to $f_{sc}\approx 40$Hz. In this work, we give a detailed description of density estimation in the solar wind using the spacecraft potential measurement from the SDP instrument on MMS. Several intervals of solar wind plasma have been processed using the methodology described, and are made available. One of the intervals is investigated in more detail and the power spectral density of the compressive fluctuations is measured from the inertial range to the sub-ion range. The morphology of the density spectra can be explained by either a cascade of Alfv\'en waves and slow waves at large scales and kinetic Alfv\'en waves at sub-ion scales or more generally by the Hall effect. Using electric field measurements the two hypotheses are discussed.

\end{abstract}



\section{Introduction} \label{sec:intro}

The solar wind is an excellent example of a turbulent plasma, which can be easily accessed for in situ plasma measurements. Plasma turbulence is characterized by disordered fluctuations in the electromagnetic fields \citep{Bale2005,Salem2012} as well as flow velocity and temperature \citep{Podesta2006,Safrankova2013,Roberts2019}. Moreover, the solar wind plasma is weakly compressive with fluctuations present in the magnitude of the magnetic field and the density \citep{Hnat2005,Roberts2017,Roberts2018}. Fluctuations are seen from the scale of the largest eddies at around $10^{6}$ km \citep{Matthaeus2005} down to the scales of the electron gyroradius at  $\sim1$ km covering seven decades in scale \citep{Kiyani2015,Verscharen2019}. Observations at scales smaller than the ion gyroradius are often limited to magnetic fluctuations as they are typically simpler to measure with the required time resolutions and sensitivities. The measurement of other plasma parameters is challenging as often apertures of instruments are pointed in a limited direction and the spin of the spacecraft allows the measurement of different azimuthal directions. The Magnetospheric MultiScale mission's (MMS) Fast Plasma Investigation (FPI) is designed so that a much higher time resolution of 0.03s is possible for electron moments using a three-dimensional distribution function. An alternative approach to derive the electron density which can be used in the solar wind is to use the spacecraft potential. Using the lower time resolution electron density data from FPI, a calibrated measurement of the electron density from the spacecraft potential can be obtained with much higher time resolution than is possible using the direct measurement. 

 To investigate the compressive fluctuations in the magnetic field either a mean magnetic field direction needs to be defined or for small fluctuations ($\delta B/B_{0} \ll 1$) the magnitude of the magnetic field can be used. At large scales, a fluid description is appropriate and magnetic and density spectra often show a Kolmogorov like spectral index of $-5/3$. This region is often termed the inertial range \citep{Tu1995,Smith2006a,Bruno2013}. When fluctuations approach ion scales, kinetic or Hall effects become important, a break is seen in the magnetic field power spectra \citep{Leamon1998,Markovskii2008,Bourouaine2012,Chen2014a,Bruno2014} which is followed by a steepening in the spectra. At the sub-ion scales the fluctuations become more compressive \citep{Kiyani2013,Roberts2017b} and the morphology of the magnetic spectrum is unclear \citep{Alexandrova2009,Alexandrova2012,Sahraoui2013}. The magnetic field spectrum is typically observed to have a spectral index close to -8/3 but can be variable due to coherent events such as parallel whistler waves or coherent structures \citep{Lion2016,Roberts2017b}.

The density spectrum has a slightly different morphology, where a flattening is often seen between the ion inertial and the ion kinetic ranges \citep{Unti1973,Neugebauer1975,Neugebauer1976,Celnikier1983}. This flattening in the spectra has been observed to have a variable range in frequency \citep{Celnikier1983,Celnikier1987,Kellogg2005,Chen2014}. Furthermore, spectra in electric fields are also variable and may differ when a monopole or dipole antenna is used. This may be due to density fluctuations affecting the monopolar measurement \citep{Kellogg2003}. The observed flattening in the density spectra has been modeled as being due to slow waves in the inertial scales which are passively cascaded in the inertial range before being heavily damped at the ion scales \citep{Harmon2005,Chandran2009,Schekochihin2009}. The smaller scales in the spectrum can be modeled as an active cascade of kinetic Alfv\'en waves. In this model, the flattening and its frequency range are related to the plasma $\beta$ (the ratio of the thermal to magnetic pressures) and are explained due to the competition between the large scale slow waves and the small scale kinetic Alfv\'en waves (KAWs). This interpretation is supported at large scales by the anti-correlation of density and magnetic field magnitude \cite[e.g.][]{Howes2011,Verscharen2017,Roberts2018}, which is a characteristic of the MHD slow-wave. Furthermore, the dispersion relations of density fluctuations found in the solar wind which show a broad range of plasma frame frequencies when compared to the trace magnetic fluctuations \citep{Roberts2017} possibly due to wave-wave interactions between Alfv\'en waves and slow waves.

In the sub-ion range, the magnetic helicity \citep[e.g.][]{He2011,Podesta2011} signature is positive when the radial component of the field is positive. The normalized magnetic and density fluctuations are also of the same order of magnitude \citep{Chen2013,Roberts2018}. Both of these observations are consistent with the KAW interpretation. Alternatively, the flattening in the density spectrum is also predicted more generally by the increasing influence of the Hall effect at small scales \citep{Narita2019,Treumann2019}. The Hall effect occurs due to the demagnetization of protons as the fluctuations in the magnetic field vary on faster timescales than the protons can follow, while electrons remain magnetized \citep{Huba2003,Kiyani2013,Narita2019,Treumann2019,Bandyopadhyay2020}. Another alternative is that this region in scale could be populated by compressible coherent structures \cite[e.g. the structures discussed by][]{Perrone2016}.  

One of the limitations in the study of the electron density spectra is that the range of scales is very limited. Often only the inertial range can be probed due to the low time resolution. To overcome this limitation remote sensing techniques have often been used \cite[e.g.][]{Woo1979,Harmon2005} or for in-situ study special plasma instruments have been developed such as FPI \citep{Pollock2016}. Alternatively, the spacecraft potential can be calibrated to give a time resolution for electron density which is equal to the electric field time resolutions \citep{Pedersen1995,Escoubet1997,Nakagawa2000,Pedersen2001,Pedersen2008}. In this study, we will use data from both the FPI instrument and the spacecraft potential from the Spin Plane double probes (SDP) instrument \citep{Lindqvist2016} to perform a new study of the electron density spectrum in the solar wind which allows frequencies deep in the sub-ion range to be investigated. Previous studies have been limited due to instrumental noise, limiting the maximum physical frequency to near 10Hz on THEMIS \citep[e.g.][]{Chen2013} or due to a time resolution of 5Hz on Cluster \citep{Yao2011,Roberts2017}. 

There are several challenges in using the spacecraft potential such as strong spin tones in the data \citep[e.g.][]{Roberts2017} and the influence of nanodust \citep[e.g.][]{Vaverka2019,Escoubet2020}. The goal of this paper is to detail the methods used for calibration of the spacecraft potential to give a measurement of the electron density, as well as the spin removal methodology which is used. This methodology is applied to 16 intervals of solar wind plasma collected in burst mode resolution from MMS which are presented in the Appendix and are made available. A further 96 intervals of fast survey mode data are also available. This data set and some calibration routines are publicly available at \burl{https://www.iwf.oeaw.ac.at/en/research/researchnbspgroups/space-plasma-physics/sc-plasma-interaction/mmsaspoc-data-analysis/}. As a demonstration of the usefulness of this approach, we also perform a detailed analysis of one interval in particular. This event will be presented in the following section as well as the methodology for calibrating the spacecraft potential to give the electron density. This will be followed by the results of the power spectral density of the electron density at sub-ion scales and will be compared to the electric and magnetic spectra. 

\section{Data/Methodology}

 \subsection{Event Overview}
\begin{figure}
    \centering
    \includegraphics[angle=-90,width=0.45\textwidth]{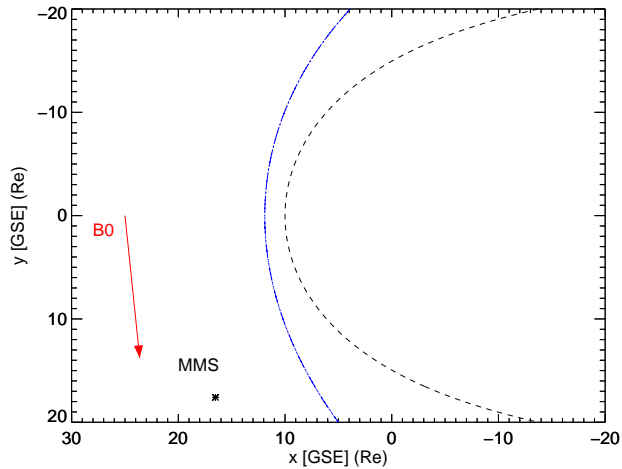}
    \caption{Location of the MMS spacecraft in GSE coordinates. The magnetopause of Earth (black dashed lines), the Bow shock (blue dot dashed lines) and the orientation of the magnetic field vector (red arrow) are also displayed.}
    \label{spacecraftLoc}
\end{figure}

The  Magnetospheirc MultiScale Mission (MMS) mission \citep{Burch2016} consists of four identical spacecraft in a tetrahedral configuration optimized for studying magnetic reconnection in the Earth's Magnetosphere. The main focus of this study is on a particular one-hour burst mode interval of slow solar wind on the 24th of November 2017 between 01:10:03-02:10:03. During this time the MMS spacecraft were located at $[x_{GSE},y_{GSE},z_{GSE}]=[16.5,17.5,6.3] R_{E}$. The subscript GSE denotes the Geocentric Solar Ecliptic coordinate system, where the $x$ component points from Earth towards the Sun, $z$ points to the North solar ecliptic. A plot of the location of the MMS spacecraft is given in figure \ref{spacecraftLoc}. The magnetopause is estimated from the model of \cite{Shue1998} and the bow shock model of \cite{Jerab2005a} is also displayed. The mean magnetic field makes a large angle with the x-direction. The location of the spacecraft and the large angle of the magnetic field with the x-direction indicate that connection with the foreshock is unlikely. This will also be validated later from the electric field spectrogram and the ion energy spectra. 

\begin{figure}[htp]
    \centering
    \includegraphics[width=0.98\textwidth]{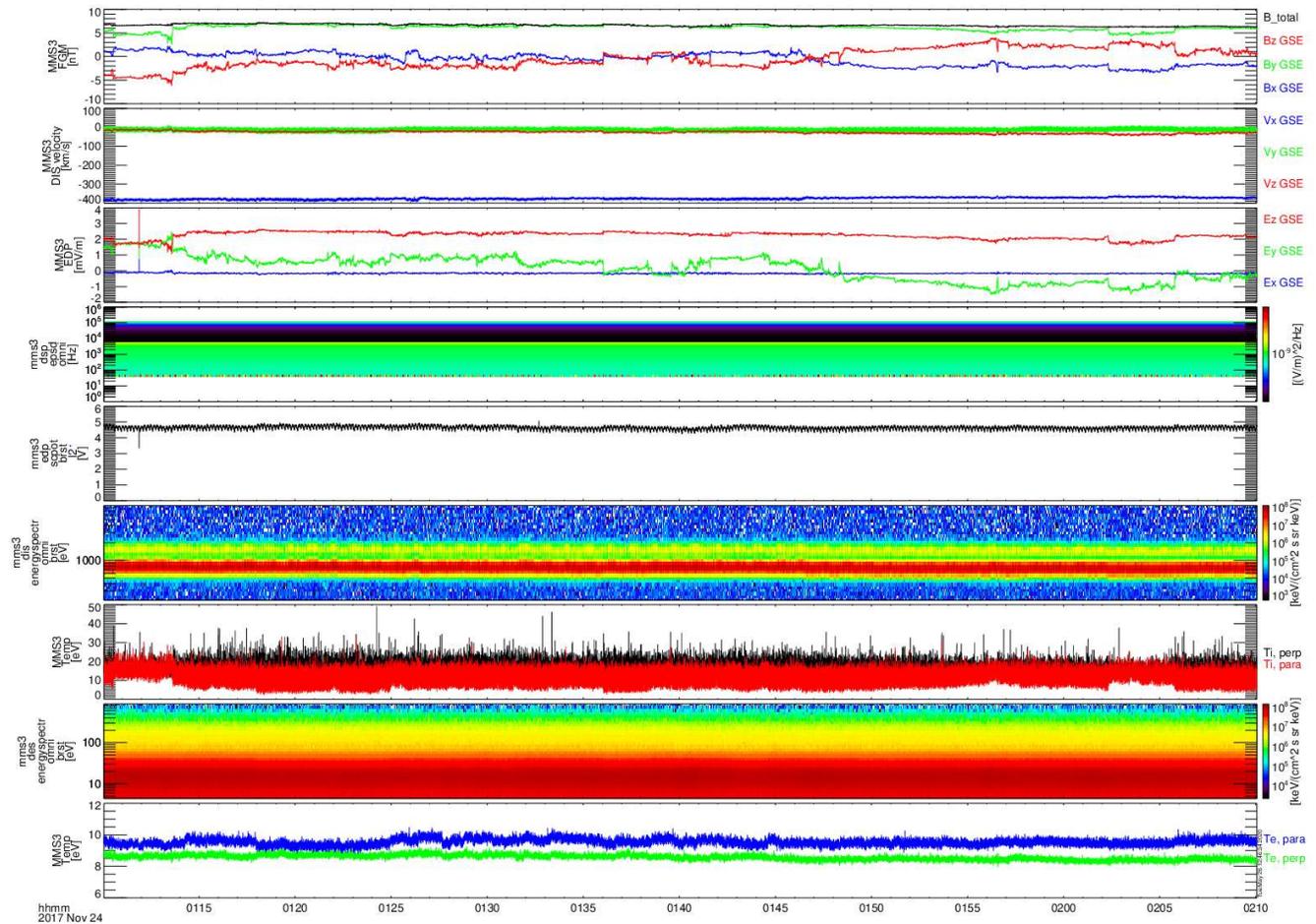}
    \caption{Measured data from the MMS3 spacecraft. The panels (from top to bottom) show the the magnetic field, the ion velocity components, the electric field, the electric field spectra, the spacecraft potential, the ion omnidirectional spectra, the ion temperature, the electron omnidirectional spectra and the electron temperatures.}
    \label{fig:data1}
\end{figure}

The magnetic field is measured by the fluxgate magnetometers (FGM) \citep{Russell2016} which have a sampling rate of 128 Hz in burst mode and sensitivity which allows the study of the magnetic fluctuations at inertial (fluid scales) and the start of the ion kinetic range before noise becomes significant at about 5Hz. The particle measurements are provided by the FPI's Dual Electron Spectrometers (DES) and Dual Ion Spectrometers (DIS) and have a sampling rate of 33Hz and 6.6Hz respectively. The spacecraft potential and the spin plane components of the electric field are obtained from the SDP \citep{Lindqvist2016} and have a sampling rate of 8.192kHz in burst mode. The third component of the electric field comes from the axial double probe (ADP) instrument \citep{Ergun2016}. A figure showing an overview of the event is shown in Fig \ref{fig:data1} and the mean plasma parameters and their respective standard deviations are presented in Tab \ref{tab:meanpars}. As there are problems for determining the ion temperature from FPI \citep{Bandyopadhyay2018} (and consequently $\beta_{i}$) the temperature from OMNI \citep{King2005} which is measured at the L1 Lagrange point and propagated to the bow shock nose is also quoted. 

Figure \ref{fig:data1} shows that there are no high energy particles, and the magnetic field makes a large angle with the radial direction with the $y$ component dominating. Furthermore, the electric field spectrogram shows no signs of high-frequency electric field waves associated with the foreshock. This is important for the study of the solar wind as a more radially pointing field may result in a connection to the foreshock, which may pollute the solar wind plasma with backstreaming particles and large amplitude wave activity \cite[e.g.][]{Turc2019}. Additionally, the electron temperature is low, and the electric fields have small amplitudes. These conditions are necessary as they can affect the determination of the electron density from the spacecraft potential \citep{Pedersen2001}. Hot electrons in the magnetosheath can cause secondary emission of electrons from the surface \citep[e.g.][]{Lai2011} with a maximum yield, which depends on the material but is typically in the range of 300-800eV. Strong electric fields or those which change abruptly can enhance the photoelectron emission from the surface \cite[e.g.][]{Torkar2017,Graham2018,Roberts2020a} causing the spacecraft potential to follow electric field fluctuations rather than the density fluctuations. In this interval the magnitude of the electric fields is small. The relevant equations and the details of the various effects on the spacecraft potential to estimate electron density will now be discussed.


\begin{table}[htp]
\centering
  \caption{Table of means and standard deviations for several plasma parameters during the one hour burst mode interval between 01:10:03-02:10:03 on 24/11/2017 taken from MMS3. The magnetic field data are from the fluxgate magnetometer and the particle data are from the Fast Plasma Investigation. Ion temperatures and $\beta_{i}$ are also shown from OMNI}
  \begin{tabular}{lccccccc}
         $B$ [nT] &  $V_{i}$ [km/s]&  $n_{e}$ [cm$^{-3}$]&$\beta_{i}$ (OMNI)&$\beta_{e}$&$T_{i}$[eV] (OMNI)&$T_{e}$[eV]\\
         $6.6\pm0.2$&$376.\pm 6.$&$8.8\pm 0.3$&$1.3\pm 0.2$&$0.7\pm 0.1$&$16.\pm 3.$&$8.7\pm 0.1$\\
         &&&$(0.4\pm 0.1)$&&$(4.7\pm     0.4)$&\
         \end{tabular}
           \label{tab:meanpars}
\end{table}

\subsection{Spacecraft Potential Calibration}
In the following sections, the methodology to obtain an electron density estimation from the spacecraft potential will be discussed. These methods will be applied to derive the electron density in 16 burst mode intervals and 96 fast survey mode intervals in total (details of which can be found in the Appendix), including the interval presented in the previous section. 

A spacecraft embedded in plasma becomes charged, and is affected by several different processes, with currents flowing to and from the spacecraft. Typically, the two dominant processes in a plasma such as the solar wind are the electron thermal current $I_{e}$ flowing to the spacecraft and the electron photocurrent flowing from the spacecraft $I_{ph}$. There can be other sources of current such as in a dense plasma the ion thermal current, or currents from the secondary emission of electrons when the electron temperature is high. There are also instruments on MMS that emit currents such as the Active Spacecraft Potential Control (ASPOC) \citep{Torkar2017} or the Electron Drift Instrument (EDI) \citep{Torbert2016} both of which are not operating in this interval. A bias current is also sent to the electric field probes, from the body of the spacecraft but is much smaller than the photoelectron or the electron thermal currents. The balance of these various currents determines the spacecraft potential. 
  
The spacecraft potential is measured on MMS from the SDP instrument which consists of four biased probes in the spacecraft spin plane which are mounted on the end of 60-meter wire booms and measure the potential difference between the probe and the spacecraft. The probe to plasma potential is stabilized by the use of a bias current which has the goal of keeping the probe to plasma potential positive and close to zero. In reality, the probes are not at zero potential with respect to the plasma, however, this would only introduce a small systematic error \cite[e.g.][]{Pedersen1995,Torkar2015}. The spacecraft potential is calculated using the four probes of the SDP instrument. If one probe is not operational due to a failure then two opposing probes are used in the calculation. The mean of the individual probe potentials give the probe to spacecraft potential, which is converted into a spacecraft to plasma potential (which we term spacecraft potential) by correcting for a boom shortening factor of 1.2 and a correction for each spacecraft one through four as $[c_{\text{MMS1}},c_{\text{MMS2}},c_{\text{MMS3}},c_{\text{MMS4}}]=[1.3,1.5,1.2,0.0]$V which are the nominal probe to plasma potentials. These were determined by investigating the regions in the electron energy spectra contaminated by photoelectrons and comparing them to the probe to spacecraft potential. The conversion of the probe to spacecraft potential $V_{psp}$ to spacecraft potential $V_{sc}$ is,
 
 \begin{equation}
     V_{sc,i}=1.2\times(-V_{psp})+c_{MMS i}
 \end{equation}
 
 where,
 
  \begin{equation}
     V_{psp}=\frac{1}{n}\sum_{i=1}^{n}V_{i}
 \end{equation}
 
where n is the number of probes used either 2 or 4.
  
In solar wind typically the electron photocurrent and the electron thermal current dominate the current balance equation such that the two can be approximately equated $I_{ph}\approx I_{e}$. The electron thermal current can be approximated, assuming a Maxwellian velocity distribution function \cite[][]{Mott-Smith1926} as:
 
 \begin{equation}
    I_{e}=-A_{\text{spac}}qn_{e}\sqrt{\frac{k_{B} T_{e}}{2 m_{e} \pi}} \left(1+\frac{q  V_{sc}}{k_{B}T_{e}}\right)
    \label{Eq2}
\end{equation}

Where $A_{\text{spac}}$ is the MMS spacecraft's approximate surface area $\sim34$m$^{2}$.

An expression for the photocurrent is found by fitting the photoelectron current (which is approximately equal thermal current) to the spacecraft potential giving the photocurve using Eq. \ref{Eq3}. 

\begin{equation}
    I_{\text{ph}}=I_{\text{phot}0}\exp{\left(-\frac{V_{sc}}{V_{\text{phot}0}}\right)}
    \label{Eq3}
\end{equation}

\begin{figure}
     \centering
     \includegraphics[angle=-90,width=0.95\textwidth]{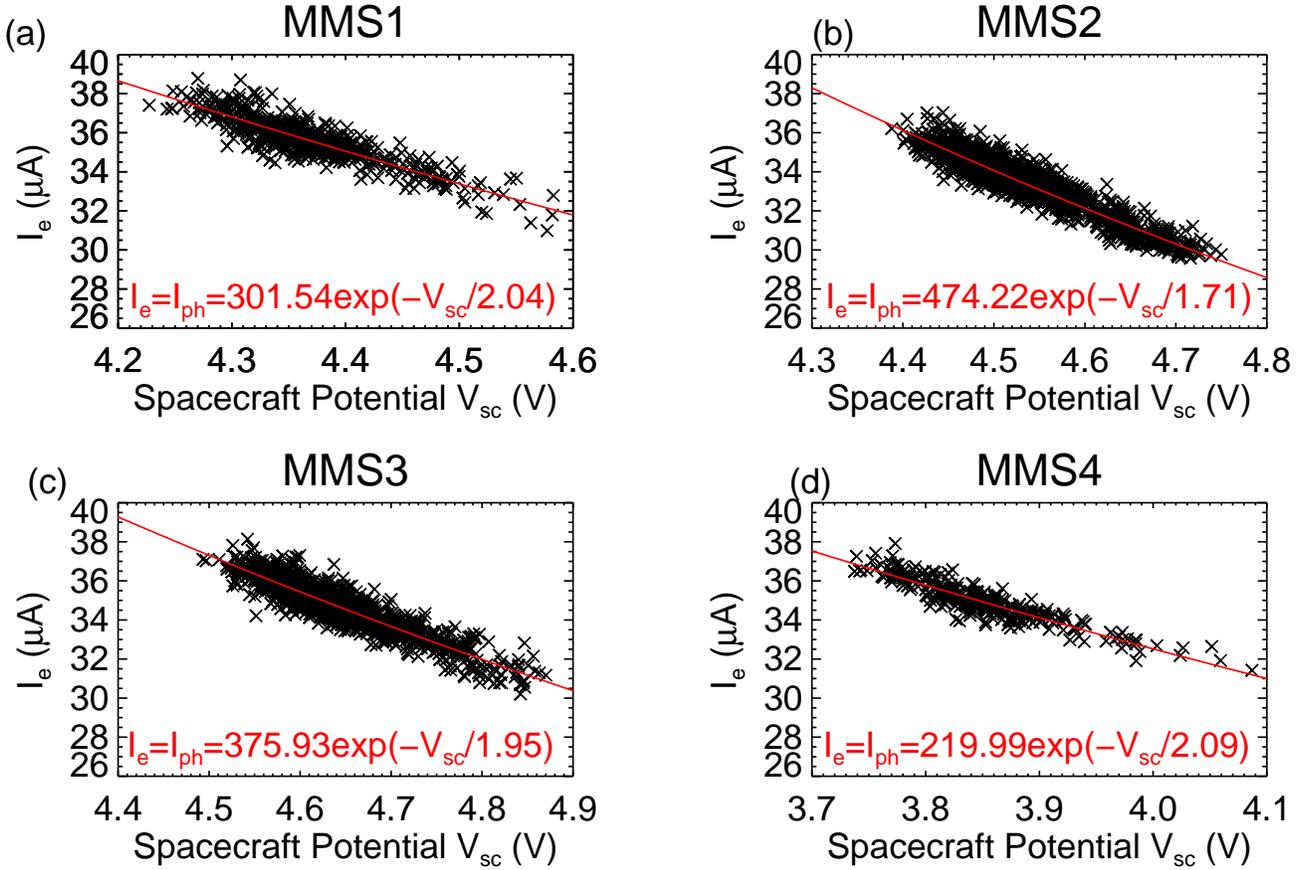}
     \caption{The photocurves and the exponential fits for all four spacecraft on the 24th of November 2017}
     \label{fig:photocurves}
 \end{figure}

The parameters $I_{\text{phot}0}$,$V_{\text{phot}0}$, are the photoelectron current and potential obtained by fitting $I_{e}$ to the spacecraft potential $V_{sc}$. The photocurve consists of several different ranges \citep{Lybekk2012,Andriopoulou2015}, where a different exponential is fitted based on the spacecraft potential. The spacecraft potential in the intervals analyzed here is always less than 5V, meaning fitting with a single exponential function is sufficient \citep{Grard1973,Andriopoulou2015}. Typically a longer time interval than the one studied is required so that a large enough range of potentials can be sampled and the relationship between the spacecraft potential and the photoelectron current can be determined effectively. Here we use the entire day of the 24th of November 2017 for calibration (the one hour of burst mode shown in Fig \ref{fig:data1} occurs on the same day) and only select times when the ion and electron density measurements from FPI are within $10\%$ of one another. This is to ensure the density measurement is accurate, as in a common plasma quasi-neutrality ($n_{e}=n_{i}$) is expected to be valid, and strong deviations from this are likely to indicate a low-quality measurement of the density. Additionally, only times when the magnitude of the DC electric field is below 10 mV/m are used for calibration. This is because strong electric fields can alter the photoelectron emission from the spacecraft causing the spacecraft potential to follow the electric field \cite[e.g.][]{Torkar2017,Graham2018a,Roberts2020a} rather than the density. The photoelectron current is modeled by performing a fit of the thermal current (Eq. \ref{Eq2}) to the spacecraft potential. Figure \ref{fig:photocurves} show the photocurves for the different spacecraft and the parameters $I_{\text{phot}0}$,$V_{\text{phot}0}$ used for calibration are indicated. MMS1-3 shows similar potentials, however, MMS4 has slightly different photoemission properties  possibly due to top ring attached to MMS4. Furthermore, there was a probe failure on MMS4 which affects the potential measurement. However, the differences in the photoemission of MMS4 were observed before the probe failure \cite[e.g.][]{Andriopoulou2018}. 

Providing that all other current sources are small, and an assumption (or a direct measurement if possible) is made about the ambient electron temperature, the spacecraft potential can be calibrated to give a measurement of the electron density (\citealt{Pedersen1995},\citealt{Escoubet1997},\citealt{Nakagawa2000},\citealt{Pedersen2001},\citealt{Pedersen2008}) given in Eq \ref{neest}.

\begin{equation}
    n_{e,SC}=\frac{1}{qA_{\text{spac}}} \sqrt{\left(\frac{2\pi m_{e}}{k_{B}T_{e}}\right)}\left(1+\frac{qV_{sc}}{k_{B}T_{e}}\right)^{-1}\left(I_{ph0}\exp\left({\frac{-V_{sc}}{V_{ph0}}}\right)\right)
    \label{neest}
\end{equation}

 As the electron thermal current varies with the electron density and the square root of the temperature the variations in the potential are more strongly affected by the electron density. Furthermore, the electron temperature is typically fairly constant in the solar wind, and was shown to be fairly constant in Fig \ref{fig:data1}.
 
\subsubsection{Spin Tone Removal}
The spacecraft potential is subject to a strong spin effect due to the sunlit area of the spacecraft changing throughout the spin, which affects the amount of photoelectron emission from the surface. This results in high power spikes in the Fourier power spectra of the spacecraft potential and the derived electron density \citep[e.g.][]{Kellogg2005,Yao2011,Chen2013,Roberts2017}. Furthermore, the spacecraft potential is measured by four probes in the spin plane of the spacecraft. As one of the probes passes the rear of the spacecraft (the front pointing at the Sun) they pass a plasma wake \cite[e.g.][]{Engwall2009}. These effects need to be removed before further analysis. This can be done by notch filtering \citep{Yao2011}, removing parts of the spectra \citep{Chen2013} subtracting harmonics, or developing an empirical model of the spacecraft charging and subtracting it \citep{Roberts2017}. The FPI instrument also suffers from some spin effects in the solar wind, in this section, we will present a method to remove such fluctuations from both the FPI density measurement and the SDP potential measurement. 

\begin{figure}[htp]
    \centering
    \includegraphics[width=0.45\textwidth]{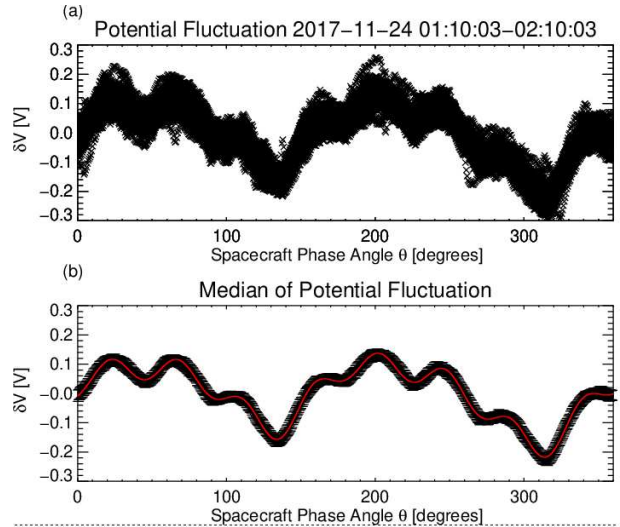}
    \caption{(a) shows the spacecraft potential data as a function of the spacecraft spin phase angle. (b) shows the median values and the errors for the median value in each bin, the red curve is the model which is fitted to the data.}
    \label{FigSDPang}
\end{figure}

We will use the same approach as \cite{Roberts2017}, where an empirical model of the spacecraft charging throughout a spin is obtained. The spacecraft potential is converted to a potential fluctuation by subtracting a suitable average. If the plasma is stable with no large changes then an average over the entire interval is sufficient. However, if there are long term trends then a moving average based on the spacecraft spin period of 20 seconds can be used.  As these are low-frequency fluctuations, the fast survey mode spacecraft potential data sampled at 32Hz can be used for obtaining the empirical model rather than the full-resolution burst mode. The spacecraft potential fluctuation is shown in Fig \ref{FigSDPang} as a function of the spacecraft spin phase angle, and a clear dependence can be seen. The potential fluctuation in Fig \ref{FigSDPang}a is binned into angle bins of 0.5 degrees and a model is derived by fitting the median values of each bin which is shown in Fig \ref{FigSDPang}b. For the MMS spacecraft, this model is based on a superposition of 20 sine waves, the model is more complicated than for Cluster studied in \cite{Roberts2017} owing to the angular octagonal shape of MMS as compared to the smoother cylindrical shape of Cluster. When the model is derived the model fluctuation is subtracted from the data removing the fluctuation due to spin leaving the other fluctuations we are interested in undisturbed.  The spin removed potential data can then be used in place of the measured potential data to obtain the electron density without spin effects. As we subtract a fluctuation, there may be uncertainties in the mean value that is subtracted. For example, if there is a long term trend the potential fluctuation in \ref{FigSDPang}a may not be centered near zero at the edges, and a mean based on a moving average should be used. Additionally, if there is a large change in the plasma conditions that perturb the potential quickly a global or even a 20-second average may not be suitable, furthermore, the wake is sensitive to the plasma conditions. Therefore this method should be applied to potentials and plasma conditions that are fairly stable. Therefore this method may appropriate when crossing large boundaries for example. 

\begin{figure}[htp]
    \centering
    \includegraphics[angle=-90,width=0.8\textwidth]{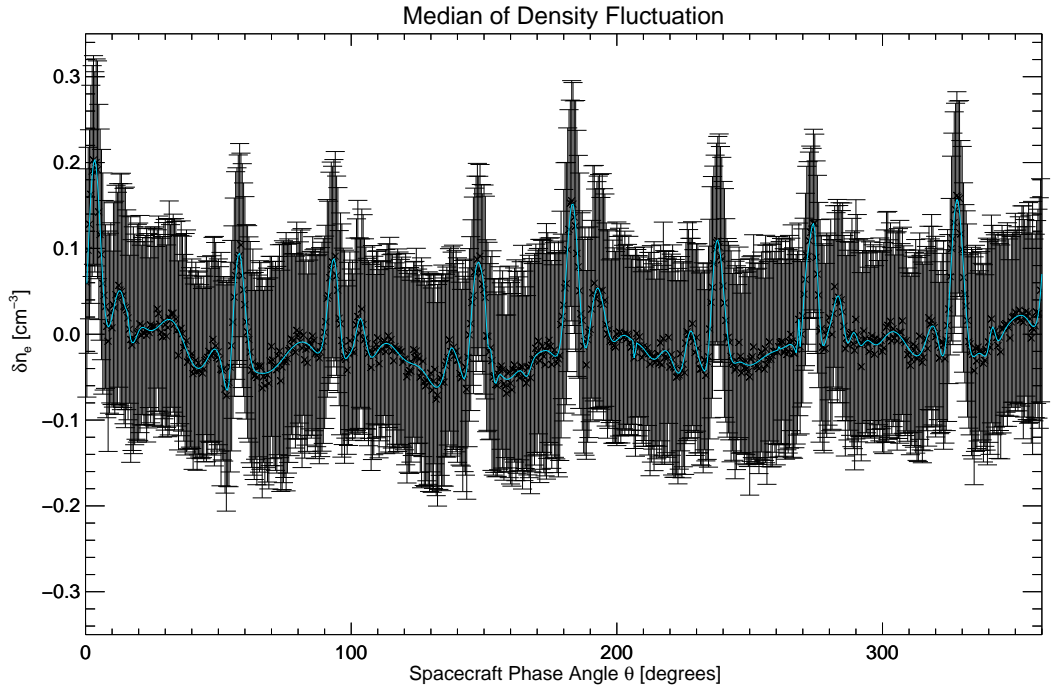}
    \caption{The median of density fluctuations in a bin of 0.5 degrees from the FPI DES measurement of electron density.}
    \label{FigFPI}
\end{figure}

The FPI data are also subject to some spin effects as the instruments are not optimized for the solar wind \citep[e.g.][]{Bandyopadhyay2018}. However, the same approach can be used to remove the spin effects from the direct measurement. Figure \ref{FigFPI} corresponds to Fig\ref{FigSDPang}b but for the FPI electron density measurement. There is also a dependence on the spacecraft spin phase angle but it is more complex. In this case, the fluctuations are fitted with a superposition of Gaussian functions and are then removed in the same way as for the spacecraft potential. 

\begin{figure}[htp]
    \centering
    \includegraphics[width=0.45\textwidth]{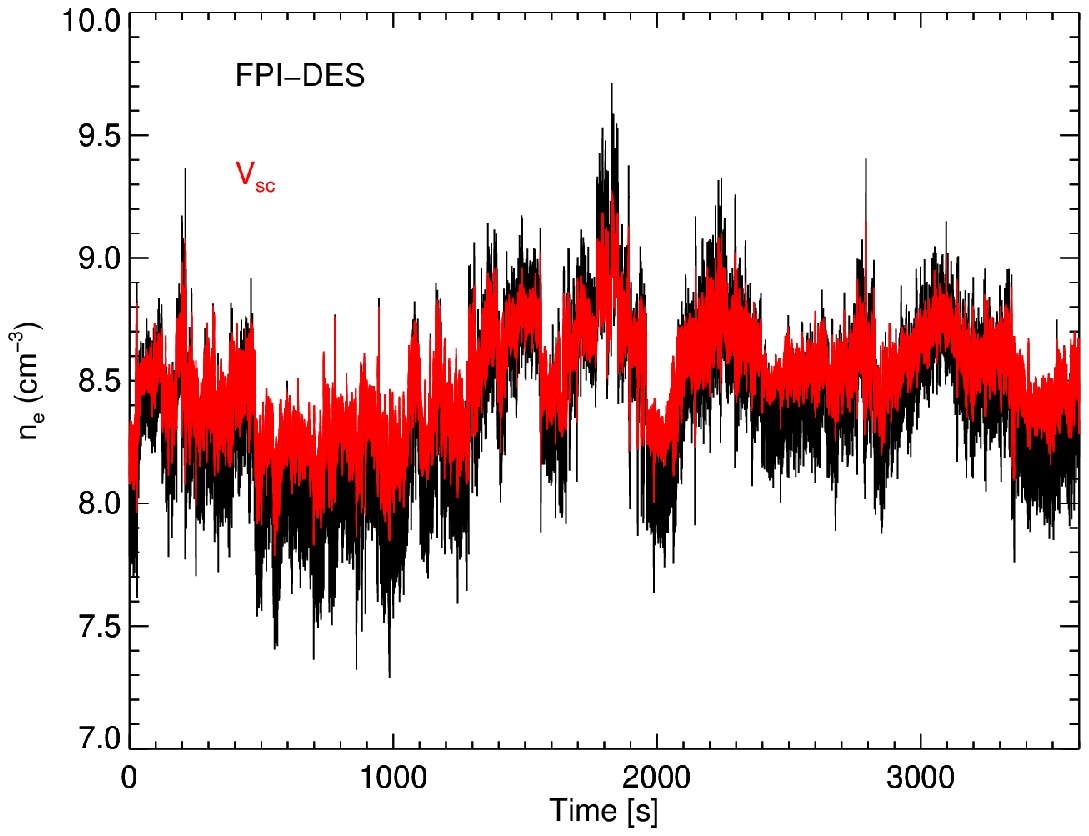}
    \caption{Comparison of the direct measurement of electron density from FPI-DES (black) and from the calibrated spacecraft potential (red).}
    \label{fig:densityComp}
\end{figure}

The Fourier Spectra of the measured electron density from FPI-DES and the spacecraft potential are shown later in Figures \ref{fig:powerspectra}a and c and the method is successful in removing the spikes. Figure \ref{fig:densityComp} shows a comparison between the FPI burst mode data and the corrected and calibrated electron density derived from the spacecraft potential on MMS3. It can be seen that there is very good agreement between both measurements. 

 \subsubsection{Dust impacts}
\begin{figure}[htp]
    \centering
    \includegraphics[width=0.8\textwidth]{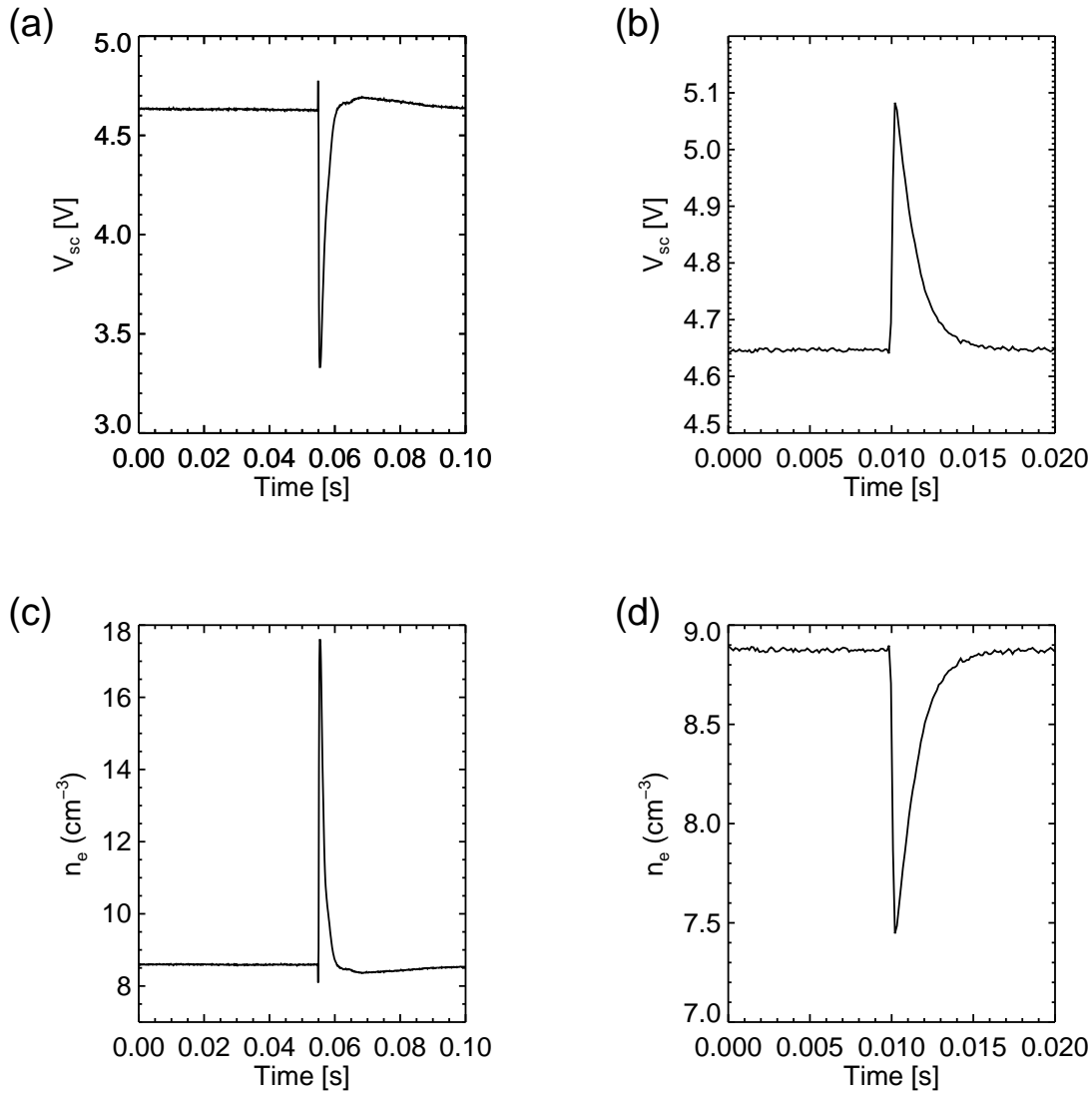}
    \caption{(a) shows an example of a dust strike event beginning at 01:11:52.729UT on MMS3.  
       (c) shows the corresponding density estimation. (b) and (d) shows a different event starting at 01:32:41.638UT where the potential increases, which has currently no satisfactory explanation.}
    \label{FigSDP}
\end{figure}

As previously mentioned dust/micrometeorites can strongly affect the spacecraft potential giving abrupt changes in the potential and the derived density. The typical profile of these is a sharp decrease in the potential followed by an increase and then an exponential tail where the spacecraft recovers to its initial state. One such example in the data interval is shown in Figure \ref{FigSDP}. The signatures shown here can also be seen in Fig\ref{fig:data1} as the large negative spike in the data near the beginning of the interval and a positive spike after 0130UT. 

The profile is seen in the potential in Figure \ref{FigSDP}a can be explained as follows; a dust impact initially causes some electrons to be lost, it then vaporizes causing a sharp increase in the plasma density near the spacecraft before the spacecraft recovers to its initial state as the spacecraft gathers electrons again  \citep[e.g.][]{Meyer-Vernet2014,Zaslavsky2015,Vaverka2017,Ye2019}. The effect on the density estimate as shown in Fig \ref{FigSDP} is extreme. It is important to remove these signatures, as it can have a large effect on some typical techniques for analyzing turbulence e.g. calculating kurtosis or other higher-order moments which are heavily influenced by outliers \citep[e.g.][]{DudokdeWit2004,Kiyani2006}.

Dust strike events have been studied on the STEREO spacecraft \citep{Malaspina2015,Oshea2017}, and were also estimated to occur on close to an hourly basis on MMS \citep{Vaverka2019}. Curiously the interval studied here is close to the peak of the Leonid meteor shower which may cause an increase in dust impacts, although definitively demonstrating a link between the meteor shower and the dust impacts seen here is not feasible. Figure \ref{FigSDP}b shows an example of a positive spike, although the interpretation of this type of signal is not clear. They are thought to be some form of dust impact. One plausible scenario is that the dust strike could occur close to the antenna \cite[e.g.][]{Malaspina2014a,Oshea2017} and that the antennas collect the electrons rather than the body of the spacecraft. Alternatively these signatures could be the result of an instrumental effect although there is no detailed explanation at present \citep{Oshea2017,Vaverka2019}. The possibility of an electron phase space hole \cite[e.g.][]{Holmes2018} is not likely judging from the fact that phase space hole events typically show a bipolar signature.

As we are interested in the density changes intrinsic to the solar wind rather than density perturbations which are caused by the interaction of the spacecraft with the surroundings (i.e. a dust particle) we will remove similar spikes by interpolating linearly. During the one hour interval, and investigating all four spacecraft 5 spikes are seen, one positive spike on MMS3 and 4 negative spikes on the other spacecraft. These are removed by linear interpolation over the time of the impacts in this paper. It is important to note that in the data provided at \burl{https://www.iwf.oeaw.ac.at/en/ffg/ffg-847969-mmsaspoc-data-analysis/}, that we do not remove dust strikes. It is up to the individual user to decide how to use the data depending on their goals.  

\subsubsection{Estimation of the Noise floor}
\begin{figure}[htp]
    \centering
    \includegraphics[width=0.95\textwidth]{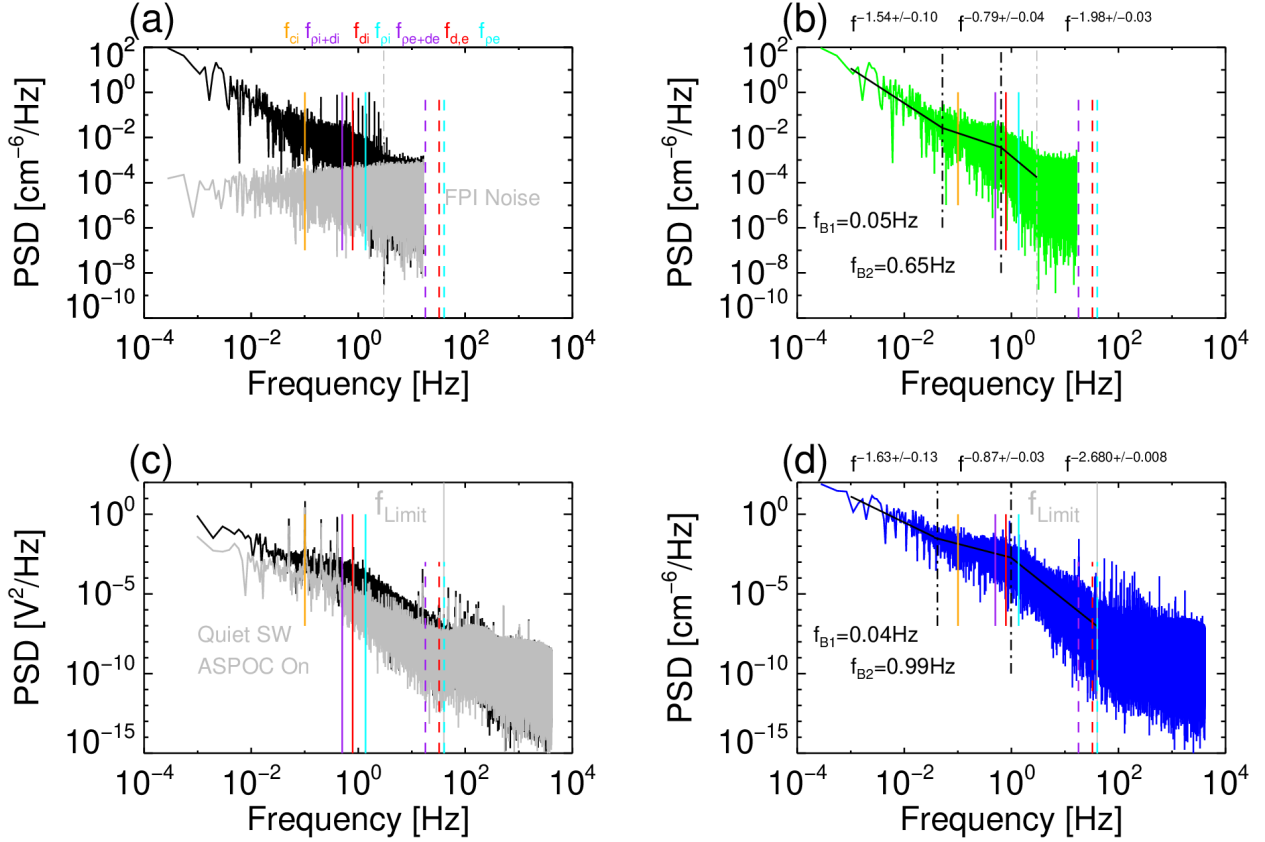}
    \caption{(a) black traces show the power spectral density of the electron density measured from FPI without spin removal. The grey trace denotes the estimated noise floor \citep{Gershman2018} the solid coloured lines denote the different characteristic scales the ion cyclotron frequency $f_{ci}$ ion Larmor radius $\rho_{i}$, the inertial length $d_{i}$, and the combined scale $f_{\rho_{i}+d_{i}}$. The dashed lines show the same for the electrons. The grey dot-dashed line at 3Hz denotes the scale where the noise becomes significant. (b) shows the spin removed electron density power spectral density and the black dot-dashed lines denote the spectral breaks. (c) shows the spectra of the measured potential data in black while the grey denotes a different interval when ASPOC is operating. (d) shows the electron density spectra obtained from the spacecraft potential.} 
    \label{fig:powerspectra}
\end{figure}

To ensure that our results are physically significant, knowledge of the noise floor is required for FPI and the SDP derived density.
In figure \ref{fig:powerspectra}, the Fourier power spectra are presented. In figures \ref{fig:powerspectra}a the measured data from FPI is shown without spin correction. A few spikes can be seen in the spectra which correspond to harmonics of the spin frequency. When the spin removal method discussed previously is applied to the FPI electron density measurement the spectra in Fig\ref{fig:powerspectra}b are obtained. The spikes in Fig \ref{fig:powerspectra}a have predominantly been removed. The estimated noise floor \citep{Gershman2018} is given in grey and noise becomes significant near 3Hz in the FPI data.

In Figure \ref{fig:powerspectra}c a 17.30-minute sub-interval between 01:10:03-01:28:33 is shown. The shorter interval is shown for direct comparison with an interval of quiet solar wind where ASPOC is on so the potential is regulated and fluctuations in the potential are smaller. The quiet interval is shown in grey. This interval on 2019/02/24 between 16:39:53-16:57:13 the electric field measurement is near the preamplifier noise at $f_{sc}>1$Hz. In reality, the noise floor is likely to be lower than the grey curve. This is because the electric field measurement is based on differences of opposing probes rather than their average therefore the electric field measurement for a component may reach the pre-amplifier noise before the spacecraft potential does. At 100Hz the spectrum flattens where white noise becomes significant, however pre-amlifier noise may affect the measurement at smaller frequencies.

\begin{figure}[htp]
    \centering
    \includegraphics[angle=90,width=0.48\textwidth]{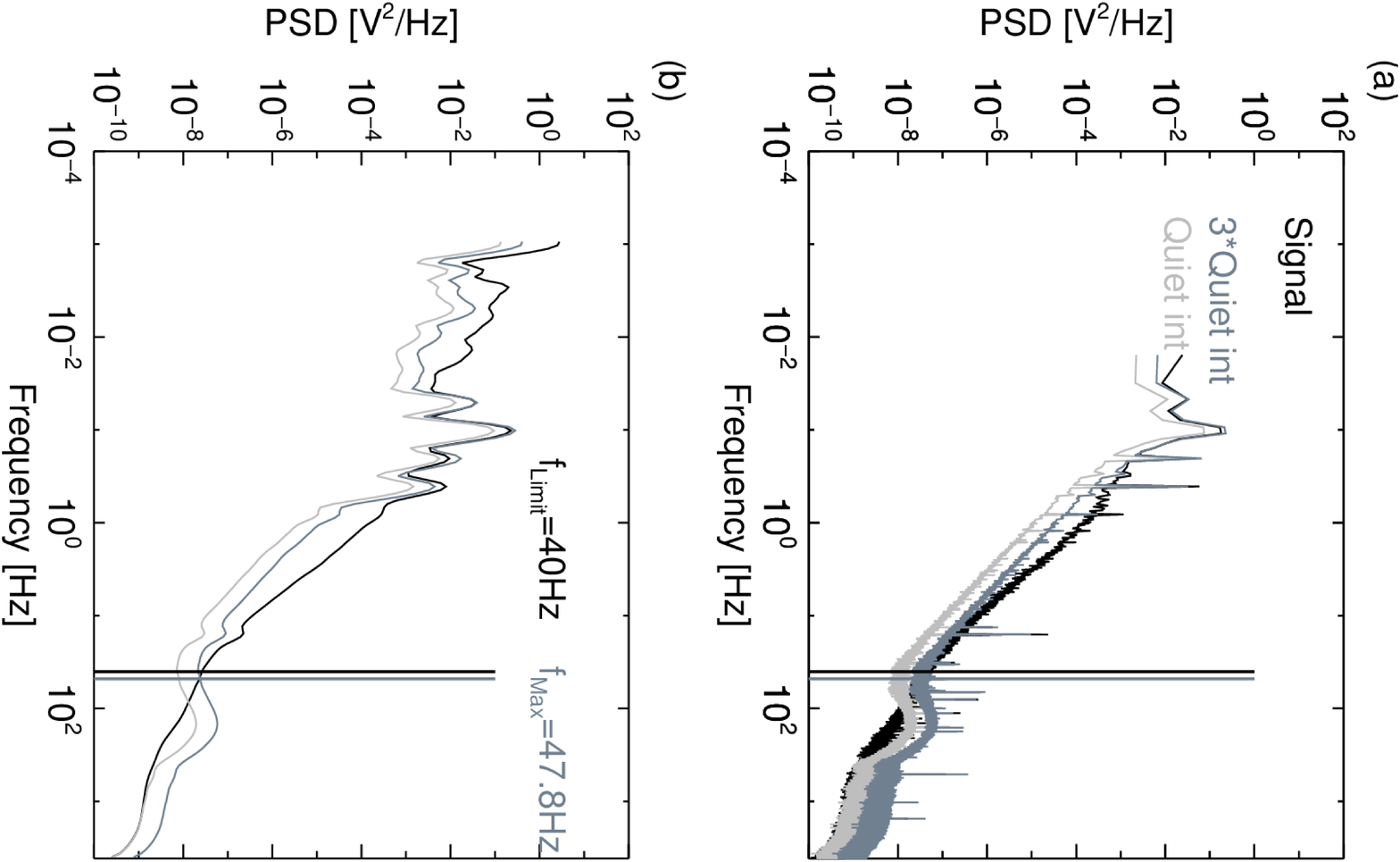}
    \caption{(a) shows the Fourier spectra from 01:10:03-01:28:33 (black) and the quiet interval where ASPOC is operating which occurs on 2019/02/24 between 16:39:53-16:57:13 (grey). The dark grey spectra denote the quiet spectra in grey which has been multiplied by 3. To better understand the noise properties the signal is separated into 64 windows and averaged. (b) shows the corresponding wavelet spectra. The vertical grey line denotes the region where the wavelet spectra are approximately equal to times the quiet signal multiplied by 3. The black line denotes where we limit our analysis to in frequency.  } 
    \label{noise}
\end{figure}

To investigate the noise floor and where the maximum physical frequency lies we use both intervals from Fig\ref{fig:powerspectra}c and calculate the Fourier spectra this time by windowing (64 windows) and averaging to reduce the variance bias. This is shown in Fig \ref{noise}a, where the signals with ASPOC on and Off are shown. We also show the power spectra of the quiet interval multiplied by 3 as the dark grey curve. Figure \ref{noise}b shows the wavelet spectra \citep{Torrence1998} have been calculated of the two signals as well as 3 times the quiet signal. The maximum physical frequency is defined as the largest frequency where the power remains above three times the noise \cite[e.g.][]{Alexandrova2010a,Roberts2017b}. For this signal, we see that this occurs at 47Hz which motivates the upper limit of 40Hz used here.

The spin removed and calibrated electron density derived from spacecraft potential is shown in Fig \ref{fig:powerspectra}d, where the spin tones prominent in Fig \ref{fig:powerspectra}c, have also been satisfactorily removed. There are some spikes at 1Hz and at higher frequencies near 16 and 32Hz and above 40Hz. These are likely to be instrumental in origin. To summarize this section; we have presented a methodology for obtaining the electron density from the spacecraft potential. This includes methods for calibration, spin removal, and an approximate estimate for the noise floor of the potential measurement. Some examples of dust strikes and inverted dust signatures have been shown. Following the methodology here we have been able to obtain a measurement of the electron density which allows the sub-ion range to be investigated.

\section{Results}

In Figures \ref{fig:powerspectra}b the FPI electron density is shown and the spectral breaks are found by fitting a straight line from either side of the break to determine the break frequency \citep{Bruno2014}. The first break in the density spectra doesn't correspond to any of the Taylor shifted ion scales, however, the second break scale is near the shifted inertial length. It is not surprising that the combined scale ($\rho_{i}+d_{i}$) associated with cyclotron resonance does not seem to link to the density spectrum here as ion cyclotron waves are not compressible. The electron density estimation from the spacecraft potential is shown in Fig \ref{fig:powerspectra}d which allows the fitting of the sub-ion range to be performed over a larger range of scales [0.983,40]Hz than for the FPI measurement [0.65,3]Hz. It is interesting to note is that there is very good agreement between both measurement methods at large scales, but at smaller scales, the spectral indices are different with the FPI measurement being significantly flatter. This is likely due to the smaller range of scales available before instrumental noise becomes significant at 3-5Hz.

\begin{figure}[htp]
    \centering
    \includegraphics[angle=90,width=0.48\textwidth]{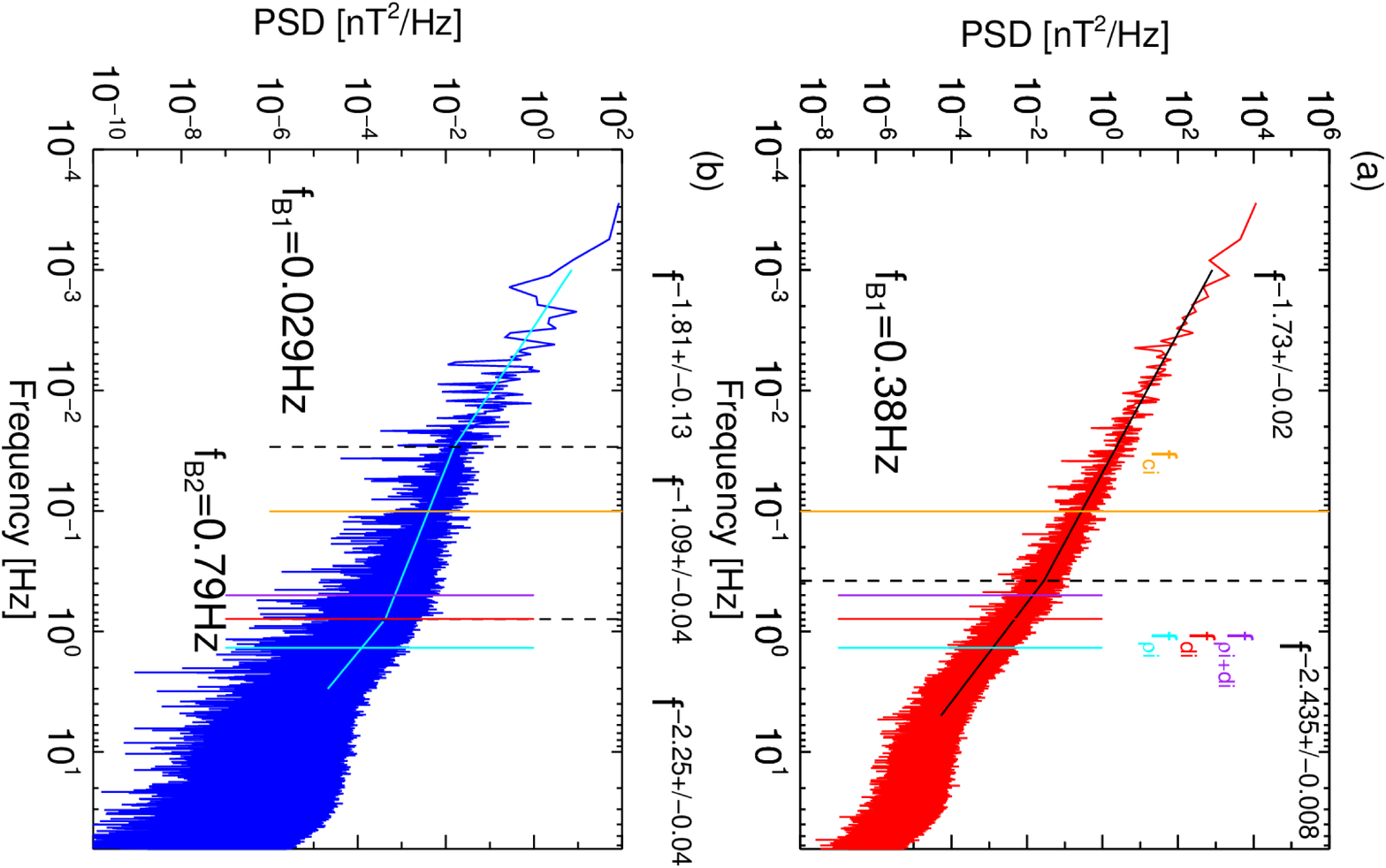}
    \caption{(a) trace magnetic power spectral density (b) magnitude power spectral density.}
    \label{fig:powerspectramagnetic}
\end{figure}

For comparison, the trace magnetic and magnitude fluctuations are also calculated from the data measured by the Fluxgate magnetometer. The magnetic spectra are shown in Fig \ref{fig:powerspectramagnetic}. It can be seen here that there is a flattening near 5Hz in the trace spectra and near 3Hz in the magnitude spectra. Unfortunately, the MMS search coil does not have the required sensitivity necessary for solar wind turbulence studies at frequencies higher than 5Hz therefore we will only use the magnetic field measurement from FGM.

The spectral break locations are found in the same manner as for the density spectra. For the trace magnetic field, the break is closest to the combined scale \citep{Bruno2014}. The error on the break in all cases is near 0.06Hz which is calculated by propagating the errors of the two linear fits. The standard deviations of the Taylor shifted scales are at most 0.07Hz. The results suggest that the magnetic field spectral break is closest to the combined scale while the density and compressible magnetic fluctuations ion scale break is closest to the ion inertial scale which is the larger of the two scales. One interpretation is that this early density break is due to kinetic slow waves which begin to be damped at lower frequencies causing the flattening seen in the density spectra. Slow waves are not the dominant source of power in the trace magnetic spectra. In the trace spectra, cyclotron resonance becomes important causing the spectral break before kinetic Alfv\'en waves become important at smaller scales leading to similar morphologies for both spectra at sub ion scales. However, this interpretation has the limitation that ion cyclotron waves are observed more often in the fast solar wind when the magnetic field is predominantly in the -x GSE (radial) direction \citep{He2011,Podesta2011} which is not the case for this interval. An alternative interpretation is that the region where the flattening is seen in the density spectra corresponds to a region where strongly compressible coherent structures are more abundant such as pressure-balanced structures. At smaller scales, the structures may have comparable powers in both the compressible and incompressible components. There is some evidence that the scale-dependent kurtosis increases with decreasing scale up to ion scales before becoming smaller again \citep{Chhiber2018,Chasapis2018}. A detailed analysis of the scale-dependent kurtosis is planned but is outside the scope of this work. 

The flattening seen in the density spectrum could also be an indication of Hall effects becoming important \citep{Narita2019,Treumann2019}. Unfortunately, the electron velocity power spectral density becomes noisy near 0.1 Hz, at lower frequencies both ion and electron velocity have a similar power spectral density which is slightly shallower than -5/3 \citep{Bandyopadhyay2018}. Multi-spacecraft increments shown in the study of \cite{Bandyopadhyay2018} do suggest at smaller scales the Hall effect is present as the velocity increments of electrons have larger power than the ions at scales of 15km. However, this is at scales far smaller than the flattening of the density spectrum observed here. An additional way to test whether it is Hall effects or the transition between the Alfv\'en and slow wave-dominated inertial range and the kinetic Alfv\'en wave kinetic range is to investigate the electric field. The ratio of the electric field to magnetic field fluctuations has been calculated previously using Cluster measurements \citep{Salem2012}. In tandem with the results of the compressibility of the fluctuations, this was interpreted as being due to kinetic Alfv\'en wave-like fluctuations.

To investigate the electric field in this interval we calculate the second-order structure functions of the electric and magnetic field. A fluctuation is defined as the difference between two time-lagged measurements $\delta \mathbf{B}(t,\tau)=\mathbf{B}(t+\tau)-\mathbf{B}(t)$ and the second-order structure-function is defined as $D^{2}(\tau)=\langle |\delta \mathbf{B}(t,\tau)|^{2}\rangle$. The mean magnetic field is predominantly in the y GSE direction $\textbf{B}_{0}=(-0.61,6.08,-0.39)$ nT, while the mean electric field direction is primarily in the z GSE direction $\textbf{E}_{0}=(-0.17,0.16,2.24)$ mV/m. It is important to note that the three orthogonal components of the electric field come from two different instruments, i.e. the x and y GSE components are calculated from the spin plane booms and the z component comes from the Axial Double Probe instrument which has lengths of approximately 60m and 12m respectively. This interval has a favorable magnetic field direction as the largest component in the electric field is measured by the instrument with the largest baseline. The SDP instrument will sample the parallel and most of one perpendicular component while the remaining perpendicular component will be measured by the ADP. However, effects due to the spacecraft wake \citep[e.g.][]{Engwall2009} will affect the SDP measurement. Effects of the wake and shadowing from the ADP make the measurement of the $E_{x}$ component difficult. In the L2 data of the electric field, these effects have been removed but some residual effects may be present. Additionally, there are spikes in the $E_{z}$ data in between two burst mode files, these are removed by linear interpolation before analysis. 

Rather than use a global magnetic field direction a local magnetic field direction will be used. This will be used for the analysis as there is increasing evidence that fluctuations in solar wind turbulence are aligned with a local magnetic field based on the size of the fluctuation rather than a global field defined by the mean over the entire time interval \cite[e.g][]{Horbury2008,Podesta2009,Chen2012,Kiyani2013}. 

\begin{figure}[htp]
    \centering
    \includegraphics[angle=-90,width=0.95\textwidth]{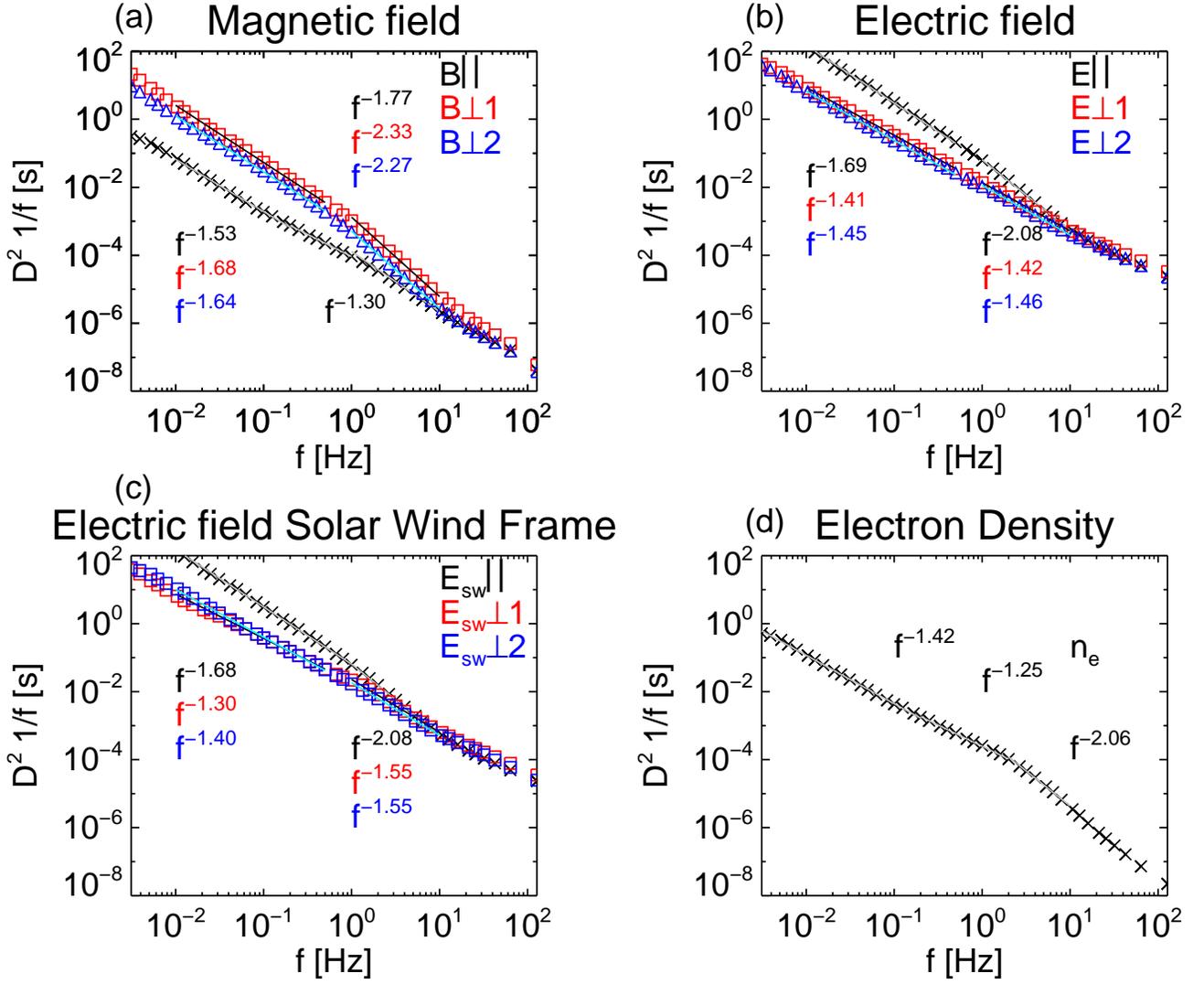}
    \caption{Second order structure functions expressed as equivalent spectra for (a) the magnetic field, (b) Spacecraft frame electric field (c) solar wind frame electric field and the (d) the electron density. }
    \label{fig2ndorder}
\end{figure}

 The coordinate system for each structure-function pair is defined as the local mean-field direction $\mathbf{B}_{\text{loc}}(t,\tau)=[\mathbf{B}(t+\tau)+\mathbf{B}(t)]/2$. The other directions are defined with respect to the unit vector of the local mean-field $\mathbf{e}_{\parallel}=\hat{\mathbf{B}}_{\text{loc}}$ where the cross product of this direction and the bulk velocity direction makes the first perpendicular direction $\mathbf{e}_{\perp 1}=\mathbf{e}_{\parallel}\times \frac{V_{sw}}{|V_{sw}|}$ and the second perpendicular direction is orthogonal to both $\mathbf{e}_{\perp 2}=\mathbf{e}_{\parallel}\times\mathbf{e}_{\perp 1}$. As the $\mathbf{e}_{\perp 2}$ is along the projection of the bulk velocity direction it can be compared to the x GSE component which is the more difficult to measure. This is further complicated by the fluctuations being very small in this component. Figure \ref{fig2ndorder} shows the second-order structure functions expressed as an equivalent spectrum, for both the electric and magnetic fields. As the electric field is not frame invariant \citep{Kellogg2006,Chen2011b,Mozer2013} the equivalent spectra are presented in the spacecraft frame (denoted subscript SC) and the solar wind frame denoted subscript (SW) which are related by the Lorentz transformation;

\begin{equation}
    \textbf{E}_{SW}=\textbf{E}_{SC}+\textbf{V}_{sw}\times\textbf{B}
\end{equation}

This transformation is done for each structure-function pair ($E(t)$ and $E(t+\tau)$) after they have been put in the local magnetic field direction, and the mean ion velocity over the entire period is used for $\mathbf{V}_{sw}$, as there are some challenges using the FPI ions point to point in the solar wind \citep{Bandyopadhyay2018}. For comparison of the amplitudes of the different spectra a normalization needs to be performed as follows; 

\begin{equation}
    \textbf{B}\rightarrow \textbf{B}/B_{0} ,
\end{equation}

\begin{equation}
    \textbf{E}\rightarrow \textbf{E}/v_{A}B_{0} ,
\end{equation}

\begin{equation}
    n_{e}\rightarrow n_{e}/n_{e0} ,
\end{equation}
 
where the subscript zeroes denote the mean over the interval.

The analysis shows that the magnetic power is dominated by the perpendicular power while the electric power is dominated by the parallel component. At large scales an MHD Alfv\'en wave does not have any associated electric field, therefore the strong parallel electric field fluctuations here are likely due to kinetic slow waves which have a large parallel component. At ion scales, the parallel electric field may be due to the KAW \citep{Narita2015}. The flattening of the electric field spectra was observed by \cite{Bale2005} and was interpreted to be due to KAW turbulence. However numerical simulations of gyrokinetic turbulence \citep{Howes2008} and Hall turbulence \citep{Matthaeus2008} both show this enhancement. The presence of large parallel electric field fluctuations suggests that Landau Damping is an important mechanism for turbulent heating. e.g. \citep{TenBarge2013,Chen2019}. The magnetic field also becomes more compressible at the start of the sub-ion range. However, noise becomes significant near 5Hz (Fig \ref{fig:powerspectramagnetic}). The electric field exhibits some flattening $f_{sc}>1$ Hz in Fig \ref{fig2ndorder} (b,c) . There is very little difference in the parallel electric field components in the two different frames, which is expected as the parallel electric field is Galilean invariant \citep{Mozer2013}. There is some difference in the $\textbf{e}_{\perp 1}$ component as it is perpendicular to both the magnetic field and approximately perpendicular to the bulk flow direction. The Lorentz transformation has the effect of flattening the $\textbf{e}_{1}$ spectra as has been observed in the statistical study in \cite{Chen2011b}. 

At large scales the parallel electric field shows a similar Kolmogorov like power law as in the magnetic field, while the perpendicular components are flatter with indices closer to -3/2 rather than -5/3 which has often been measured for the velocity fluctuations in the solar wind \citep{Podesta2006,Bandyopadhyay2018} and in the magnetosheath \cite{Roberts2019}. At sub-ion scales, it is difficult to make any firm conclusions as noise becomes significant. As the electric field measurement is based on taking the difference in the potential between two probes, the noise is likely to be a larger problem for the electric field than the spacecraft potential. To contrast with the potential measurement, the measurement is based on an average of the four probes meaning that the signal to noise ratio will be larger for the potential measurement allowing frequencies up to 40Hz to be resolved when compared to a few Hz for the electric field measurement. With the data available it is difficult to discriminate between the KAW scenario and the Hall scenario. However, in this region, the spectrum does steepen slightly for all components especially in the solar wind frame. This steepening is predicted for KAW turbulence before the spectrum flattens at higher wavenubmers \citep{Narita2020}, while Hall Turbulence would exhibit only a flattening \citep{Narita2019}. Although it should be noted that there is only a short range of scales where the steepening is seen and it is difficult to determine conclusively between both scenarios. Furthermore, the flattening predicted at kinetic scales in the electric field is difficult to distinguish from noise \citep[e.g.][]{Alexandrova2013a}. Perhaps more subtly both effects are present and contribute to the observations. The strong parallel electric field fluctuations do however support the dissipation of energy through Landau damping. 

\begin{figure}[htp]
    \centering
    \includegraphics[width=0.45\textwidth]{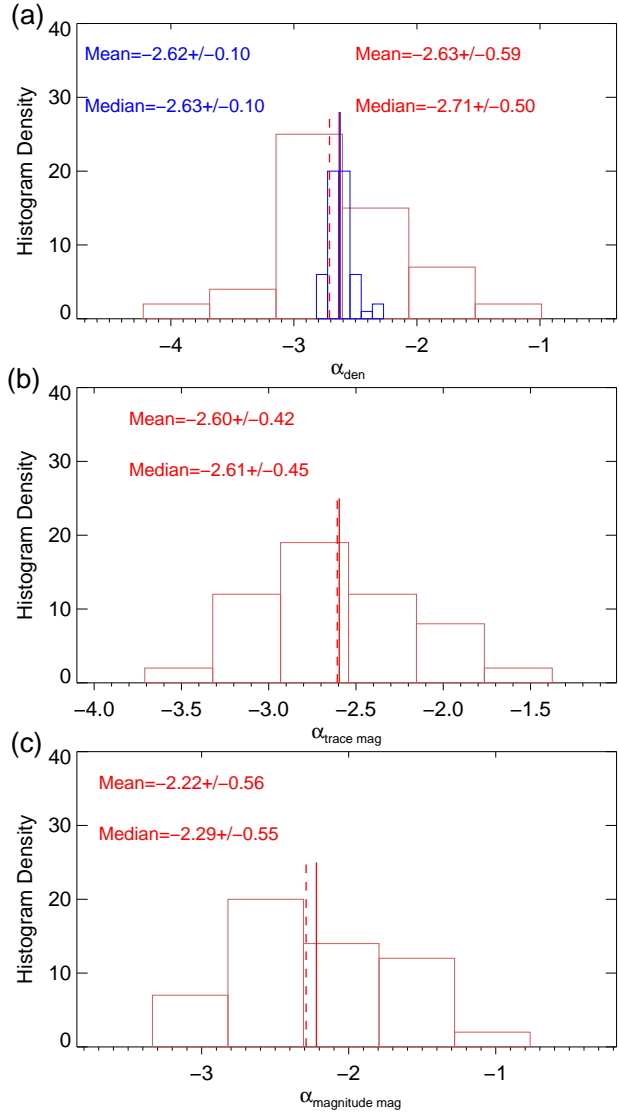}
    \caption{(a) shows the histogram of the spectral indices of the density power spectral density fitted between 1 and 3Hz (red) and between 1 and 40Hz in blue. (b) shows the fitting of the trace magnetic field spectra fitted between 1 and 3 Hz (c) shows the same for the power spectral density of the magnitude. The solid and dashed lines denote the mean and the median values of the spectral indices}
    \label{fig:histgrams}
\end{figure}

At sub-ion scales, the relationship between the spectral indices of the magnetic and the density spectra is investigated by performing by fitting the slopes of the power spectral density between the frequencies of 1Hz and 3Hz. The results are presented as histograms in Figure \ref{fig:histgrams}. As it is the ion kinetic scales that concern us rather than the spectral break locations we can split the time interval into smaller intervals of 64 seconds ($2^{19}$ data points for the spacecraft potential and $2^{13}$ for the magnetic field). The power spectral density is estimated using a Hanning window with no overlap, and are averaged over 7 windows to reduce the variance. From the one hour interval, there is a total of 55 spectra which we use to investigate the spectral index in the sub-ion range.

The mean and median values of the spectral indices for the density are close to -2.6 which compare well with magnetic field measurements of this range given in Figure \ref{fig:powerspectra}b and are consistent with previous studies of the magnetic spectra \citep{Smith2006a,Alexandrova2012,Sahraoui2013}. However there is a large spread in the values, as has been seen in \cite{Smith2006a} for the dissipation range in a statistical study of the magnetic field spectra using the ACE spacecraft containing various plasma conditions. For example, a change in the orientation of the magnetic field could cause the spectral index to change \cite{Horbury2008,Wicks2011,Roberts2017,Roberts2019}. As the interval is only one hour and the dissipation range indices show high variability we interpret this as being a result of the limited frequency range where we can fit both magnetic and density spectra.

To test this hypothesis we perform the same fittings over the frequency range 1-40Hz for the density spectra which is shown in the blue values in figure \ref{fig:histgrams}a and the spread in the values is significantly limited, however, the mean values are consistent over both frequency ranges. Therefore it is important to fit spectra over the largest range possible, otherwise one might conclude that there is a large variability in the spectrum where a fitting over a larger frequency range reveals that this is not correct. It is also noted that the spectra of the magnitude are flatter than the trace and the density spectra. This is due to the noise becoming an issue at higher frequencies causing an artificial flattening in the spectra similar to what is seen in the FPI density spectra in Figure \ref{fig:powerspectra}.

\begin{figure}[htp]
    \centering
    \includegraphics[angle=90,width=0.45\textwidth]{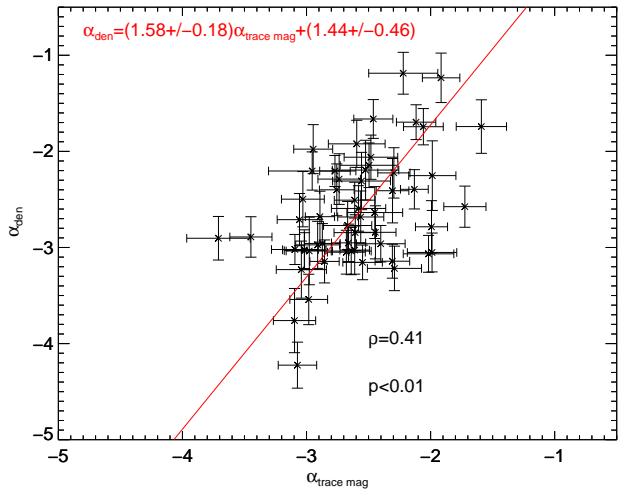}
    \caption{Comparison of the spectral indices of the density and the trace magnetic field spectra in the range [1,3]Hz. The error bars are from the residuals of fitting a straight line to the log of frequency to the log of power.}
    \label{fig:indcomp}
\end{figure}

In Figure \ref{fig:indcomp} we compare our values with the spectral indices which are obtained from the magnetic field spectra for the same intervals in Figures \ref{fig:histgrams}. For direct comparison, we fit them over the shorter frequency range between 1-3 Hz. The error bars denote the residuals from the least-squares fitting of log frequency to log power. There is a moderate correlation between the two values of the spectral indices of with a Spearman coefficient of 0.41 and a $p<0.01$ between the two spectral indices \citep{Chen2013}.  This result is expected as at these scales there is evidence that the fluctuations become more compressible towards isotropy in this range \cite{Kiyani2013,Roberts2017a}. This was interpreted by \cite{Kiyani2013} to be a characteristic of kinetic Alfv\'en waves as linear theory predicts KAWs become strongly compressible in the sub-ion range.

\section{Conclusions}
In this paper, a methodology for using the spacecraft potential for the estimation of electron densities in the solar wind has been provided with MMS. In terms of time resolution, it can outperform the direct measurement of the electron density from FPI giving an additional decade in scale. There are several caveats present in the data such as strong spin effects due to the sunlit area of the spacecraft varying throughout a spin, dust strikes, and some instrumental spikes at higher frequencies. A method is provided to remove spin tones, and the same principle can be used to remove the spin effects in the FPI data. An example of a dust strike and a possible inverted dust like signature has been shown in the interval analyzed. In the data set provided at \burl{https://www.iwf.oeaw.ac.at/en/ffg/ffg-847969-mmsaspoc-data-analysis/} these strikes are not removed. It is up to the individual user to decide how to treat them or if they are the subject of their investigations themselves. However a user should be aware especially if they are interested in higher-order statistics \cite[e.g.][]{Chasapis2017,Chhiber2018} as one example. It has been demonstrated that (1) sub-ion scale compressive turbulence can be investigated with the spacecraft potential on MMS (2) Calculating the slope of the power spectrum in a limited region of frequency space such as we have done with FPI in Fig \ref{fig:powerspectra}b can be misleading in terms of the spectral index and in terms of the variability of the spectral index (3) The morphology of the density spectrum shows similarities to the trace magnetic spectra at the inertial range and the kinetic range, however in between these two ranges there is a flattening (4). The spectral breaks occur at different frequencies for the density and the trace magnetic fluctuations. Finally, the signatures of the electric field support the interpretation that compressive fluctuations at large scales are slow wave-like and that smaller-scale fluctuations are likely to be kinetic Alfv\'en wave-like. The fact that the parallel electric field is dominant suggests that there is ample energy that is available for Landau damping.  In this interval we are unable to determine whether there is an 'electron dissipation range' \citep{Sahraoui2010a,Sahraoui2013} or an exponential cutoff \citep{Alexandrova2009,Alexandrova2012} at the electron scales as the potential data for this interval becomes noisy at these scales. However, it might be possible to perform measurements with the potential at these scales in the future should there be a strongly compressive interval where the signal to noise ratio is much larger.

\acknowledgments
\sloppy
The datasets analyzed for this study can be found in the MMS science data archive \burl{https://lasp.colorado.edu/mms/sdc/public/ } . This includes the original spacecraft potential which is calibrated to the electron density. The calibrated electron density data as well as several useful codes to analyze spacecraft potential data are available at \burl{https://www.iwf.oeaw.ac.at/en/research/researchnbspgroups/space-plasma-physics/sc-plasma-interaction/mmsaspoc-data-analysis/}. Analysis of the spacecraft potential data at IWF is supported by Austrian FFG projects ASAP15/873685. R.N. was supported by Austrian FWF projects I2016-N20. Z. V. was supported by the Austrian FWF projects P28764-N27. OWR acknowledges helpful discussions with Eva Schunova-Lilly and Ali Varsani.

%

\appendix
\section{Solar wind intervals analyzed}

\begin{table}[h]
  \caption{Table of the mean and standard deviations of solar wind burst mode intervals where the density calibration has been performed. Plasma $\beta$ is calculated by reasmpling the magnetic field data onto the FPI time tags, to avoid large values near the edges the first and last seconds are removed from the calculation. Consequently should there be magnetic holes the standard deviation will be very large. The one minute resolution OMNI data are used for the calculation of $\beta$. This is only calculated for times when  $T_{i}$,$B$, and $n_{i}$ are all available.}
  \label{IntervalTableBurst}

\scriptsize
  \begin{tabularx}{\textheight}{llcccccccccc}
  \centering
         Date&Time&$B$ [nT] &$V_{i}$ [km/s]&$n_{e}$ [cm$^{-3}$]&$\beta_{i}$ (OMNI)&$\beta_{e}$&$T_{i}$[eV] (OMNI)&$T_{e}$[eV]\\
         2016-12-06&11:37:34-11:44:03&$7.6\pm0.4$&$350.\pm 9.$&$18\pm 2$&$1.6\pm 0.3$&$1.4\pm 0.2$&$15\pm 2$&$11.1\pm 0.3$\\
         &&&&&($0.5 \pm 0.1$)&&($2.9\pm 0.1$)&\\
         2016-12-31&08:19:54-08:25:03&$6.3\pm0.1$&$307.\pm 8.$&$20.5\pm 0.2$&$2.4\pm 0.4$&$2.4\pm 0.1$&$12\pm 2$&$11.7\pm 0.2$\\
         &&&&&$(0.5\pm0.1)$&&$(1.4\pm0.2)$&&&&\\
         2017-11-24&01:10:03-02:10:02&$6.6\pm0.2$&$376.\pm 6.$&$8.8\pm 0.3$&$1.3\pm 0.2$&$0.7\pm 0.1$&$16.\pm 3.$&$8.7\pm 0.1$\\
         &&&&&$(0.4\pm    0.1)$&&$(4.7\pm     0.4)$&&&&\\
         2017-11-26&21:09:03-22:09:02&$7.7\pm0.2$&$334.\pm 5.$&$11.0\pm 0.9$&$1.1\pm 0.2$&$0.9\pm 0.1$&$12.\pm 2.$&$12.2\pm 0.3$\\
         &&&&&$(0.3\pm 0.1)$&&$(3.8\pm 0.4)$&&&&\\
         2017-11-28&19:45:23-20:16:02&$7.3\pm0.1$&$410.\pm 10.$&$7.7\pm 0.3$&$1.9\pm 0.5$&$0.8\pm 0.1$&$31.\pm 9.$&$13.8\pm 0.5$\\
         &&&&& $    (0.2\pm   0.1)$&&$(2.6\pm     0.2)$&&&&\\
         2017-12-02&14:50:03-15:50:02&$4.6\pm0.3$&$396.\pm 6.$&$4.4\pm 0.2$&$2.0\pm 0.6$&$0.7\pm 0.1$&$22.\pm 6.$&$8.5\pm 0.2$\\
         &&&&& $(0.3\pm 0.1)$&&$(3.3\pm 0.4)$&&&&\\
         2017-12-18&14:40:03-15:34:52&$3.3\pm0.2$&$620.\pm 10.$&$2.1\pm 0.1$&$8.\pm 3.$&$1.6\pm 0.3$&$100.\pm 33.$&$20.\pm 2.$\\
        &&&&& $(1.2\pm0.4)$&&$(18\pm 2)$&&&&\\
        2018-01-08&04:00:03-05:00:02&$6.1\pm0.2$&$298.\pm 5.$&$13.1\pm 0.5$&$1.4\pm 0.3$&$1.2\pm 0.1$&$9.\pm 2.$&$8.0\pm 0.4$\\
         &&&&& $(0.2\pm 0.1)$&&$(1.5\pm 0.1)$&&&&\\
        2018-01-28&02:55:03-03:10:52&$3.4\pm0.2$&$391.\pm 3.$&$6.8\pm 0.2$&$3.6\pm 0.7$&$2.0\pm 0.3$&$15.\pm 3.$&$8.8\pm 0.1$\\
         &&&&& $(1.6\pm 0.2)$&&$(5.0\pm0.5)$&&&&\\
        2019-11-30&16:00:33-17:00:32&$4.0\pm0.4$&$393.\pm 8.$&$11.6\pm 0.6$&$6.\pm 2.$&$3.1\pm 0.7$&$19.\pm 3.$&$10.5\pm 0.5$\\
        &&&&& $(1.4\pm 0.2)$&&$(4.4\pm0.6)$&&&&\\
        2019-12-04&00:00:53-01:00:52&$3.6\pm0.7$&$314.\pm 6.$&$16.8\pm 0.9$&$8.\pm 4.$&$6.\pm 3.$&$14.\pm 3.$&$9.9\pm 0.4$\\
        &&&&& $(2\pm 1)$&&$(1.6\pm 0.2)$&&&&\\
        2019-12-05&03:07:33-03:57:32&$3.0\pm0.2$&$304.\pm 4.$&$10.4\pm 0.4$&$5.\pm 2.$&$3.4\pm 0.5$&$11.\pm 3.$&$7.4\pm 0.2$\\
        &&&&& $(0.7\pm0.2)$&&$(1.5\pm0.1)$&&&&\\
        2019-12-19&03:00:03-04:00:02&$5.8\pm0.6$&$521.\pm 4.$&$5.5\pm 0.7$&$4.\pm 2.$&$0.7\pm 0.3$&$45\pm 12$&$9.4\pm 0.4$\\
        &&&&& $(1.4\pm0.7)$&&$(18.0\pm      1.2)$&&&&\\
        2019-12-22&19:18:43-19:55:02&$5.0\pm0.1$&$347.\pm 4.$&$7.2\pm 0.2$&$1.7\pm 0.3$&$1.2\pm 0.1$&$15\pm 3.$&$10.5\pm 0.2$\\
        &&&&& $(0.4\pm  0.1)$&&$(3.5\pm     0.5)$&&&&\\
        2020-01-02&05:50:13-07:00:02&$3.4\pm0.2$&$314.\pm 5.$&$9.3\pm 0.4$&$4.\pm 1.$&$2.3\pm 0.4$&$12.\pm 2.$&$6.7\pm 0.1$\\
        &&&&& $(0.7\pm 0.1)$&&$(2.4\pm     0.1)$&&&&\\
        2020-01-31&01:30:03-01:45:02&$7.1\pm0.1$&$448.\pm 4.$&$3.7\pm 0.2$&$0.9\pm 0.1$&$0.4\pm 0.1$&$30.\pm 4.$&$13.7\pm 0.3$\\
        &&&&& $(0.3\pm 0.1)$&&$(12.\pm 1.)$&&&&\\
    \end{tabularx}
  \end{table}
  \pagebreak
  \vspace*{5cm}
  \begin{table}[h]
  \caption{Table of the mean and standard deviations of solar wind fast survey mode intervals where the density calibration has been performed 2016-2017.}
  \label{IntervalTable20162017}

\scriptsize
  \begin{tabularx}{\textheight}{llcccccccccc}
  \centering
         Date&Time&$B$ [nT] &$V_{i}$ [km/s]&$n_{e}$ [cm$^{-3}$]&$\beta_{i}$ (OMNI)&$\beta_{e}$&$T_{i}$[eV] (OMNI)&$T_{e}$[eV]\\
         2016-12-06&11:20:01-11:59:55&$7.7\pm0.8$&$349.\pm 5.$&$17.\pm 2.$&$2.6\pm 0.9$&$1.5\pm 0.4$&$21.\pm 1.$&$12.6\pm 0.6$\\
         &&&&&($0.4\pm0.2$)&&($2.5\pm0.2 $)&\\
         2016-12-31&08:00:00-11:14:56&$8.\pm2.$&$310.\pm 10.$&$23.\pm 3.$&$5.\pm 8.$&$3.\pm 2.$&$20.\pm 20.$&$14.\pm 2.$\\
         &&&&&($0.4\pm0.2$)&&($1.5\pm0.2 $)&\\
         2017-11-14&20:00:04-22:59:59&$5.0\pm0.8$&$390.\pm 10.$&$6.3\pm 0.7$&$4.\pm 2.$&$1.7\pm 0.6$&$34.\pm 4.$&$15.4\pm 0.8$\\
         &&&&&($0.5\pm0.3$)&&($4.\pm1. $)&\\
         2017-11-15&12:00:00-13:55:21&$9.5\pm0.7$&$394.\pm 6.$&$8.\pm 2.$&$1.\pm 1.$&$0.3\pm 0.1$&$32.\pm 9.$&$6.3\pm 0.7$\\
         &&&&&($0.12\pm0.01$)&&($2.5\pm0.8 $)&\\
         2017-11-17&15:10:00-16:29:57&$4.1\pm0.2$&$418.\pm 6.$&$4.9\pm 0.2$&$5.2\pm 0.7$&$1.8\pm 0.2$&$44.\pm 4.$&$14.\pm 1.$\\
         &&&&&($0.7\pm0.2$)&&($5.5\pm0.8 $)&\\
         2017-11-18&06:30:00-09:39:58&$3.8\pm0.2$&$400.\pm 10.$&$5.6\pm 0.4$&$8.\pm 3.$&$1.9\pm 0.3$&$50\pm 20.$&$12.\pm 1.$\\
         &&&&&($0.6\pm0.3$)&&($4.0\pm1.0 $)&\\
         2017-11-20&11:30:04-13:29:59&$4.0\pm0.1$&$325.\pm 3.$&$8.7\pm 0.4$&$5.8\pm 0.8$&$3.8\pm 0.4$&$27.\pm 4.$&$17.9\pm 0.8$\\
         &&&&&($0.29\pm0.04$)&&($1.4\pm0.1 $)&\\    
         2017-11-21&03:15:04-06:14:55&$12.\pm2.$&$490.\pm 20.$&$12.\pm 2$&$4.\pm 2.$&$0.5\pm 0.2$&$130.\pm 10.$&$15.3\pm 0.9$\\
        &&&&&($1.0\pm0.6$)&&($28.\pm6. $)&\\
        2017-11-23&22:36:36-23:59:56&$6.2\pm0.3$&$380.\pm 7.$&$7.8\pm 0.5$&$3.\pm 1.$&$0.7\pm 0.1$&$30.\pm 10.$&$8.9\pm 0.1$\\
        &&&&&($0.33\pm0.06$)&&($3.9\pm0.4 $)&\\
        2017-11-24&00:00:00-02:39:23&$6.6\pm0.2$&$373.\pm 9.$&$8.6\pm 0.5$&$3.2\pm 0.9$&$0.7\pm 0.1$&$40.\pm 10.$&$8.9\pm 0.1$\\
        &&&&&($0.32\pm0.06$)&&($4.2\pm 0.5 $)&\\
        2017-11-26&19:00:00-20:39:59&$4.2\pm0.9$&$301.\pm 8.$&$11.\pm 1.$&$5.\pm 4.$&$2.4\pm 1.0$&$20.\pm 12.$&$9.0\pm 0.7$\\
        &&&&&($1.0\pm0.5$)&&($3.3\pm 0.7 $)&\\
        2017-11-28&19:30:00-21:19:59&$6.6\pm0.6$&$412.\pm 6.$&$9.\pm 1.$&$3.1\pm 0.8$&$1.3\pm 0.4$&$39.\pm 5.$&$15.5\pm 0.6$\\
        &&&&&($0.3\pm0.2$)&&($3.2\pm 0.6 $)&\\
        2017-11-29&15:10:00-17:49:59&$7.5\pm0.4$&$383.\pm 7.$&$11.\pm 1.$&$4.\pm 1.$&$1.1\pm 0.2$&$56.\pm 7.$&$14.0\pm 0.6$\\
        &&&&&($0.5\pm0.2$)&&($6.\pm 2. $)&\\
        2017-12-01&16:40:00-19:21:59&$4.2\pm0.7$&$436.\pm 4.$&$5.3\pm 0.3$&$6.\pm 4.$&$2.\pm 1.$&$44\pm 4.$&$15.3\pm 0.7$\\
        &&&&&($2.\pm1.$)&&($8.\pm1. $)&\\
        2017-12-02&14:10:00-17:09:59&$4.3\pm0.7$&$396.\pm 3.$&$4.6\pm 0.3$&$3.\pm 1.$&$0.9\pm 0.4$&$30.\pm 3.$&$8.4\pm 0.4$\\
        &&&&&($0.3\pm0.2$)&&($2.9\pm0.4 $)&\\
        2017-12-04&10:30:00-13:59:59&$4.0\pm0.8$&$316.\pm 2.$&$12.\pm 2.$&$8.\pm 8.$&$5.\pm 4.$&$20.\pm 2.$&$11.2\pm 0.9$\\
        &&&&&($2.\pm2.$)&&($2.3\pm0.3 $)&\\
        2017-12-05&07:40:00-09:39:59&$10.\pm2.$&$490.\pm 30.$&$11.\pm 2.$&$5.\pm 3.$&$0.6\pm 0.4$&$110.\pm 20$&$13.4\pm 0.8$\\
        &&&&&($0.9\pm0.7$)&&($19.\pm5. $)&\\
        2017-12-08&03:05:00-04:04:59&$3.1\pm0.3$&$451.\pm 4.$&$3.3\pm 0.2$&$6.\pm 2.$&$2.2\pm 0.5$&$46.\pm 7.$&$16.\pm 2.$\\
        &&&&&($0.9\pm0.3$)&&($7.\pm1. $)&\\
        2017-12-11&00:00:03-00:39:59&$3.5\pm0.2$&$330.\pm 3.$&$9.1\pm 0.4$&$5.2\pm 0.6$&$3.4\pm 0.5$&$17.\pm 1.$&$12.\pm 1.$\\
        &&&&&($0.48\pm0.06$)&&($1.8\pm0.2 $)&\\
        2017-12-13&16:00:00-20:09:59&$4.1\pm0.4$&$432.\pm 7.$&$4.3\pm 0.3$&$7.\pm 2.$&$1.7\pm 0.4$&$60\pm 10.$&$15.\pm 2.$\\
        &&&&&($0.6\pm0.2$)&&($6.\pm1. $)&\\
        2017-12-15&18:40:00-19:39:59&$2.5\pm0.6$&$358.\pm 5.$&$9.7\pm 0.9$&$20.0\pm 8.$&$8.\pm 3.$&$27.\pm 2.$&$10.6\pm 0.5$\\
        &&&&&($1.7\pm0.4$)&&($1.5\pm0.2$)&\\
        2017-12-22&05:40:00-10:39:59&$3.9\pm0.5$&$324.\pm 6.$&$6.6\pm 0.8$&$3.\pm 1.$&$1.7\pm 0.6$&$17.\pm 2.$&$9.3\pm 0.4$\\
        &&&&&($0.4\pm0.2$)&&($2.5\pm0.7 $)&\\
        2017-12-27&18:40:00-23:59:59&$8.\pm1.$&$410.\pm 16.$&$10.\pm 1.$&$5.\pm 4.$&$1.0\pm 0.6$&$52.\pm 5.$&$11.2\pm 0.9$\\
        &&&&&($0.6\pm0.3$)&&($8.\pm2. $)&\\
        2017-12-30&13:30:00-18:59:59&$4.2\pm0.8$&$362.\pm 6.$&$6.7\pm 0.6$&$7.\pm 4.$&$2.\pm 1.$&$41.\pm 8.$&$10.8\pm 0.8$\\
&&&&&($0.9\pm0.8$)&&($4\pm1. $)&\\

    \end{tabularx}
  \end{table}
  \pagebreak
   \vspace*{2cm}
   
  \begin{table}[h]
  \caption{Table of the mean and standard deviations of solar wind fast survey mode intervals where the density calibration has been performed 2018.}
  \label{IntervalTable2018}

\scriptsize
  \begin{tabularx}{\textheight}{llcccccccccc}
  \centering
         Date&Time&$B$ [nT] &$V_{i}$ [km/s]&$n_{e}$ [cm$^{-3}$]&$\beta_{i}$ (OMNI)&$\beta_{e}$&$T_{i}$[eV] (OMNI)&$T_{e}$[eV]\\

2018-01-02&08:30:00-14:09:59&$5.\pm2.$&$420.\pm 10.$&$7.0\pm 0.9$&$11.\pm 40.$&$2.\pm 8.$&$60.\pm 10.$&$13.\pm 1.$\\
&&&&&($2.\pm4.$)&&($7.\pm 1. $)&\\
2018-01-05&09:00:00-10:39:59&$2.5\pm0.3$&$359.\pm 3.$&$7.4\pm 0.7$&$9.\pm 3.$&$4.\pm 1.$&$19.\pm 1.$&$8.1\pm 0.7$\\
&&&&&($1.2\pm0.6$)&&($2.4\pm0.6 $)&\\
2018-01-11&02:00:00-03:29:59&$3.1\pm0.3$&$404.\pm 2.$&$5.0\pm 0.4$&$6.\pm 1.$&$2.8\pm 0.6$&$30.\pm 2.$&$13.\pm 2.$\\
&&&&&($1.2\pm0.5$)&&($4.9\pm0.6 $)&\\
2018-01-13&22:00:00-03:39:59&$8.\pm2.$&$560.\pm 30.$&$13.\pm 1.$&$10.\pm 40.$&$2.\pm 5.$&$120.\pm 10.$&$15.\pm 1.$\\
&&&&&($2.\pm2.$)&&($32.\pm6. $)&\\
2018-01-19&06:00:00-11:29:59&$8.6\pm0.6$&$360.\pm 9.$&$12.\pm 1.$&$2.3\pm 0.5$&$1.0\pm 0.2$&$35.\pm 7.$&$15.\pm 1.$\\
&&&&&($0.15\pm0.04$)&&($2.2\pm0.5 $)&\\
2018-01-22&06:40:00-12:09:59&$5.\pm1.$&$510.\pm 10.$&$3.3\pm 0.6$&$5.\pm 3.$&$0.7\pm 0.3$&$100.\pm 20.$&$13.\pm 1.$\\
&&&&&($0.7\pm0.6$)&&($11.\pm4. $)&\\
2018-01-25&02:50:00-08:19:59&$7.6\pm0.9$&$385.\pm 7.$&$11.\pm 2.$&$4.\pm 3.$&$1.1\pm 0.6$&$48.\pm 5.$&$13.\pm 1.$\\
&&&&&($0.9\pm0.6$)&&($10.\pm2. $)&\\
2018-01-30&21:50:00-23:09:59&$2.7\pm0.5$&$329.\pm 2.$&$12.8\pm 0.9$&$13.\pm 5.$&$7.\pm 2.$&$16.\pm 1.$&$8.6\pm 0.2$\\
&&&&&($3.\pm2.$)&&($1.7\pm0.2 $)&\\
2018-01-31&00:00:00-02:39:59&$2.3\pm0.2$&$327.\pm 2.$&$13.2\pm 0.5$&$20.\pm 10.$&$9.\pm 7.$&$16.1\pm 0.7$&$8.2\pm 0.3$\\
&&&&&($1.7\pm0.2$)&&($1.6\pm0.1 $)&\\
2018-02-10&01:10:00-06:19:59&$5.\pm2.$&$332.\pm 6.$&$12.\pm 2.$&$6.\pm 4.$&$3.\pm 2.$&$20.\pm 2.$&$11.\pm 2.$\\
&&&&&($0.7\pm0.4$)&&($2.4\pm0.3 $)&\\
2018-02-19&14:00:00-14:59:59&$7.1\pm0.3$&$540.\pm 10.$&$2.9\pm 0.1$&$2.4\pm 0.3$&$0.3\pm 0.1$&$100.\pm 10.$&$12.2\pm 0.5$\\
&&&&&($0.38\pm0.06$)&&($17.\pm3. $)&\\
2018-02-19&16:00:00-17:29:59&$6.5\pm0.4$&$537.\pm 9.$&$3.0\pm 0.3$&$3.2\pm 0.6$&$0.5\pm 0.1$&$110.\pm 10.$&$16.\pm 2.$\\
&&&&&($1.\pm1.$)&&($13.\pm5. $)&\\
2018-02-22&10:50:00-13:59:59&$6.3\pm0.8$&$393.\pm 9.$&$12.\pm 1.$&$6.\pm 6.$&$2.\pm 2.$&$44.\pm 7.$&$13.1\pm 0.9$\\
&&&&&($0.7\pm0.6$)&&($5.9\pm0.6 $)&\\
2018-03-04&19:00:00-21:39:59&$3.4\pm0.6$&$347.\pm 7.$&$6.6\pm 0.8$&$6.\pm 5.$&$3.\pm 2.$&$24.\pm 3.$&$11.\pm 1.$\\
&&&&&($1.2\pm0.6$)&&($3.4\pm0.7 $)&\\
2018-03-08&15:30:00-16:59:59&$5.9\pm0.3$&$331.\pm 3.$&$8.2\pm 0.4$&$2.2\pm 0.9$&$0.9\pm 0.1$&$22.\pm 7.$&$9.1\pm 0.6$\\
&&&&&($0.36\pm0.06$)&&($3.6\pm0.4 $)&\\
2018-03-(16-17)&22:30:00-02:59:59&$8.\pm1.$&$450.\pm 10.$&$8.\pm 2.$&$4.\pm 2.$&$0.6\pm 0.2$&$80.\pm 10.$&$12.\pm 1.$\\
&&&&&($0.7\pm0.3$)&&($14.\pm3. $)&\\
2018-03-25&12:00:00-15:59:59&$4.6\pm0.7$&$447.\pm 7.$&$7.2\pm 0.6$&$10.\pm 10.$&$2.\pm 3.$&$50.\pm 10.$&$10.5\pm 0.9$\\
&&&&&($1.\pm1.$)&&($7.\pm1. $)&\\
2018-04-(07-08)&21:30:00-01:59:59&$3.2\pm0.4$&$367.\pm 2.$&$5.0\pm 0.4$&$5.\pm 2.$&$2.6\pm 0.8$&$27.\pm 3.$&$12.1\pm 0.7$\\
&&&&&($0.6\pm0.2$)&&($3.0\pm0.6 $)&\\

    \end{tabularx}
  \end{table}
  \pagebreak
  \vspace*{1cm}
    \begin{table}[h]
  \caption{Table of the mean and standard deviations of solar wind fast survey mode intervals where the density calibration has been performed 2019.}
  \label{IntervalTable2019}

\scriptsize
  \begin{tabularx}{\textheight}{llcccccccccc}
  \centering
         Date&Time&$B$ [nT] &$V_{i}$ [km/s]&$n_{e}$ [cm$^{-3}$]&$\beta_{i}$ (OMNI)&$\beta_{e}$&$T_{i}$[eV] (OMNI)&$T_{e}$[eV]\\
  2019-03-(06-07)&20:30:00-05:59:59&$5.7\pm0.8$&$380.\pm 10.$&$5.3\pm 0.9$&$2.\pm 2.$&$0.7\pm 0.4$&$25.\pm 7.$&$10.2\pm 0.9$\\
&&&&&($0.5\pm0.5$)&&($5.\pm2.$)&\\
2019-03-10&09:30:00-15:59:59&$3.5\pm0.4$&$377.\pm 8.$&$2.8\pm 0.5$&$4.\pm 1.$&$1.4\pm 0.3$&$40.\pm 10.$&$15.2\pm 0.7$\\
&&&&&($0.4\pm0.1$)&&($6.\pm2. $)&\\
2019-03-14&04:45:00-06:39:59&$4.7\pm0.2$&$340.\pm 3.$&$6.5\pm 0.5$&$2.1\pm 0.3$&$1.2\pm 0.1$&$18.0\pm 1.0$&$10.3\pm 0.1$\\
&&&&&($0.25\pm0.05$)&&($2.0\pm0.3 $)&\\
2019-03-17&14:15:00-17:09:59&$7.6\pm0.2$&$383.\pm 6.$&$5.2\pm 0.5$&$0.9\pm 0.2$&$0.3\pm 0.1$&$23.\pm 2.$&$7.1\pm 0.4$\\
&&&&&($0.09\pm0.02$)&&($2.2\pm0.3 $)&\\
2019-03-19&20:07:00-21:06:59&$5.0\pm0.3$&$400.\pm 2.$&$8.0\pm 0.4$&$3.8\pm 0.6$&$2.3\pm 0.4$&$29.\pm 2.$&$17.\pm 1.$\\
&&&&&($0.9\pm0.2$)&&($7.6\pm0.5 $)&\\
2019-03-31&20:10:00-21:19:59&$3.7\pm0.3$&$402.\pm 4.$&$4.8\pm 0.2$&$5.\pm 1.$&$1.4\pm 0.2$&$36.\pm 9.$&$9.5\pm 0.3$\\
&&&&&($1.0\pm0.4$)&&($7.0\pm0.4 $)&\\
2019-04-07&18:35:00-20:59:59&$3.2\pm0.2$&$377.\pm 3.$&$4.5\pm 0.3$&$3.8\pm 0.7$&$1.6\pm 0.3$&$21.\pm 2.$&$9.0\pm 0.5$\\
&&&&&($0.5\pm0.1$)&&($2.6\pm0.5 $)&\\
2019-04-18&06:45:00-09:29:59&$5.0\pm0.4$&$308.\pm 4.$&$5.2\pm 0.6$&$1.4\pm 0.3$&$0.8\pm 0.2$&$16.\pm 1.$&$9.6\pm 0.6$\\
&&&&&($0.31\pm0.08$)&&($3.0\pm0.7 $)&\\
2019-11-09&18:00:00-19:29:59&$3.1\pm0.1$&$324.\pm 2.$&$6.3\pm 0.2$&$6.4\pm 0.9$&$3.3\pm 0.4$&$23.\pm 2.0$&$12.3\pm 0.5$\\
&&&&&($0.9\pm0.2$)&&($2.8\pm0.4 $)&\\
2019-11-10&11:10:00-13:59:59&$4.1\pm0.3$&$301.\pm 6.$&$5.3\pm 0.4$&$4.1\pm 0.9$&$1.6\pm 0.3$&$31.\pm 5.$&$12.0\pm 0.4$\\
&&&&&($0.15\pm0.04$)&&($1.0\pm0.2 $)&\\
2019-11-(13-14)&23:50:00-01:59:59&$4.4\pm0.5$&$327.\pm 4.$&$4.9\pm 0.9$&$3.\pm 2.$&$1.3\pm 0.6$&$25.\pm 4.$&$11.\pm 1.$\\
&&&&&($0.4\pm0.3$)&&($2.1\pm0.6 $)&\\
2019-11-17&10:50:00-11:29:59&$4.2\pm0.1$&$429.\pm 3.$&$3.3\pm 0.1$&$3.3\pm 0.4$&$1.7\pm 0.1$&$42.\pm 4.$&$22.5\pm 0.5$\\
&&&&&($0.36\pm0.08$)&&($4.9\pm0.9 $)&\\
2019-11-21&21:30:00-22:27:59&$6.4\pm0.6$&$591.\pm 6.$&$5.8\pm 0.4$&$5.\pm 1.$&$1.0\pm 0.3$&$80.\pm 4$&$17.\pm 2.$\\
&&&&&($2.\pm1.$)&&($25.\pm5. $)&\\
2019-11-23&17:30:00-18:09:59&$7.3\pm0.3$&$433.\pm 6.$&$4.3\pm 0.2$&$1.4\pm 0.2$&$0.5\pm 0.1$&$43.\pm 4.$&$16.\pm 2.$\\
&&&&&($0.4\pm0.1$)&&($8.\pm2. $)&\\
2019-11-30&13:30:00-19:09:59&$4.1\pm0.8$&$391.\pm 6.$&$10.7\pm 0.9$&$8.\pm 4.$&$4.\pm 2.$&$26.\pm 1.$&$12.9\pm 0.9$\\
&&&&&($1.8\pm0.7$)&&($5.\pm2. $)&\\
2019-12-04&00:00:00-06:59:59&$4.\pm1.$&$311.\pm 4.$&$14.\pm 2.$&$8.\pm 8.$&$5.\pm 5.$&$17.\pm 2.$&$12.2\pm 0.7$\\
&&&&&($1.4\pm0.9$)&&($2.0\pm0.3 $)&\\
2019-12-05&03:00:00-05:57:59&$2.6\pm0.4$&$307.\pm 5.$&$9.\pm 1.$&$9.\pm 4.$&$5.\pm 2.$&$15.\pm 1.$&$8.\pm 1.$\\
&&&&&($1.\pm2.$)&&($1.7\pm0.2 $)&\\
2019-12-16&18:40:00-19:49:59&$3.9\pm0.4$&$340.\pm 3.$&$3.9\pm 0.2$&$3.8\pm 0.9$&$1.9\pm 0.4$&$35.\pm 4.$&$18.3\pm 0.8$\\
&&&&&($0.36\pm0.08$)&&($3.5\pm0.9 $)&\\
2019-12-18&00:00:01-04:59:59&$5.\pm2.$&$360.\pm 20.$&$15.\pm 5.$&$10.\pm 70.$&$8.\pm 45.$&$23.\pm 3.$&$13.7\pm 0.9$\\
&&&&&($2.\pm3.$)&&($4.\pm 2. $)&\\
2019-12-22&19:05:00-19:54:59&$5.0\pm0.1$&$346.\pm 2.$&$7.1\pm 0.3$&$2.2\pm 0.2$&$1.2\pm 0.1$&$19.\pm 1.$&$10.7\pm 0.3$\\
&&&&&($0.5\pm0.5$)&&($3.4\pm0.7 $)&\\
2019-12-29&15:30:00-17:39:59&$4.5\pm0.5$&$292.\pm 3.$&$10.\pm 1.$&$2.9\pm 0.8$&$1.5\pm 0.5$&$13.\pm 1.$&$6.7\pm 0.4$\\
&&&&&($0.3\pm0.1$)&&($1.3\pm0.2 $)&\\

    \end{tabularx}
  \end{table}
  \pagebreak
  \vspace*{1cm}
  
      \begin{table}[htp]
  \caption{Table of the mean and standard deviations of solar wind fast survey mode intervals where the density calibration has been performed 2020.}
  \label{IntervalTable2020}
  
  \scriptsize
  \begin{tabularx}{\textheight}{llcccccccccc}
  \centering
         Date&Time&$B$ [nT] &$V_{i}$ [km/s]&$n_{e}$ [cm$^{-3}$]&$\beta_{i}$ (OMNI)&$\beta_{e}$&$T_{i}$[eV] (OMNI)&$T_{e}$[eV]\\
2020-01-02&05:20:00-07:19:59&$3.4\pm0.3$&$312.\pm 3.$&$9.2\pm 0.5$&$6.\pm 3.$&$2.3\pm 0.7$&$18.\pm 4.$&$7.0\pm 0.4$\\
&&&&&($0.9\pm0.3$)&&($2.4\pm0.2 $)&\\
2020-01-04&12:00:00-15:59:59&$6.\pm1.$&$371.\pm 7.$&$3.4\pm 0.4$&$1.9\pm 0.5$&$0.7\pm 0.3$&$40.\pm 8.$&$14.\pm 1.$\\
&&&&&($0.13\pm0.05$)&&($2.0\pm0.4 $)&\\

2020-01-10&09:40:00-12:19:59&$6.1\pm0.9$&$493.\pm 4.$&$4.0\pm 0.4$&$2.5\pm 0.7$&$0.8\pm 0.2$&$56.\pm 5.$&$17.\pm 1.$\\
&&&&&($0.6\pm0.2$)&&($9.\pm1. $)&\\
2020-01-12&17:20:00-20:09:59&$2.8\pm0.2$&$380.\pm 10.$&$3.0\pm 0.1$&$5.\pm 1.$&$2.0\pm 0.5$&$30.\pm 2.$&$12.9\pm 0.6$\\
&&&&&($0.8\pm0.3$)&&($5.\pm1. $)&\\
2020-01-19&20:00:00-21:59:59&$1.0\pm0.3$&$304.\pm 3.$&$7.9\pm 0.5$&$100.\pm 300.$&$100\pm 100.0$&$17.\pm 2.$&$7.3\pm 0.5$\\
&&&&&($4.\pm2.$)&&($1.0\pm0.2 $)&\\
2020-01-23&10:05:00-11:14:59&$4.0\pm0.1$&$335.\pm 6.$&$5.2\pm 0.3$&$2.6\pm 0.3$&$1.6\pm 0.2$&$20.\pm 2.$&$12.2\pm 0.9$\\
&&&&&($0.39\pm0.09$)&&($2.9\pm0.4 $)&\\
2020-01-26&03:10:00-06:59:59&$2.2\pm0.9$&$296.\pm 6.$&$10.\pm 2.$&$30.\pm 60.$&$20.\pm 40.$&$13.6\pm 0.8$&$7.8\pm 0.4$\\
&&&&&($4.\pm4.$)&&($2.2\pm0.7 $)&\\
2020-01-27&04:00:00-06:59:59&$3.5\pm0.5$&$308.\pm 3.$&$9.\pm 1.$&$5.\pm 2.$&$3.\pm 1.$&$15.\pm 1.$&$9.0\pm 0.6$\\
&&&&&($0.8\pm0.9$)&&($2.1\pm0.5 $)&\\
2020-01-(29-31)&13:00:00-03:44:59&$6.\pm1.$&$410.\pm 20.$&$5.\pm 1.$&$4.\pm 50$&$1.\pm 11.$&$50.\pm 10.$&$11.\pm 2.$\\
&&&&&($1.\pm3.$)&&($9.\pm3. $)&\\
2020-02-02&17:00:00-18:59:59&$4.0\pm0.3$&$395.\pm 5.$&$4.3\pm 0.2$&$3.3\pm 0.5$&$1.3\pm 0.2$&$29.\pm 2.$&$11.9\pm 0.3$\\
&&&&&($0.7\pm0.2$)&&($6.0\pm0.9 $)&\\
2020-02-03&05:00:00-07:29:59&$3.2\pm0.3$&$378.\pm 2.$&$4.2\pm 0.3$&$4.2\pm 0.9$&$1.6\pm 0.3$&$25.\pm 2.$&$9.4\pm 0.4$\\
&&&&&($0.5\pm0.2$)&&($2.7\pm0.5 $)&\\

2020-02-06&00:00:00-05:59:59&$7.\pm2.$&$360.\pm 20.$&$6.\pm 3.$&$5.\pm 200$&$1.\pm 40.$&$40.\pm 10.$&$12.\pm 1.$\\
&&&&&($0.4\pm0.3$)&&($6.\pm2. $)&\\

2020-02-(06-07)&14:00:00-01:59:59&$7.7\pm0.9$&$540.\pm 40.$&$6.6\pm 0.8$&$5.\pm 3.$&$0.7\pm 0.3$&$110.\pm 20.$&$14.\pm 2.$\\
&&&&&($0.9\pm0.5$)&&($21.\pm4. $)&\\

2020-02-08&13:30:00-16:29:59&$2.9\pm0.3$&$476.\pm 4.$&$2.5\pm 0.2$&$8.\pm 3.$&$2.1\pm 0.8$&$68.\pm 7.$&$17.2\pm 0.8$\\
&&&&&($0.9\pm0.5$)&&($6.\pm2.$)&\\

2020-02-09&21:10:00-22:29:59&$2.9\pm0.4$&$424.\pm 5.$&$3.7\pm 0.2$&$7.\pm 3.$&$2.3\pm 0.8$&$35.\pm 3.$&$11.9\pm 0.5$\\
&&&&&($0.6\pm0.2$)&&($4.2\pm0.6 $)&\\

2020-02-12&01:00:00-03:59:59&$3.6\pm0.7$&$353.\pm 5.$&$8.\pm 1.$&$7.\pm 5.$&$3.\pm 2.$&$24.\pm 2.$&$12.\pm 1.$\\
&&&&&($1.\pm1.$)&&($2.7\pm0.7 $)&\\

2020-02-19&02:10:00-04:59:59&$8.\pm2$&$410.\pm 10.$&$10.\pm 2.$&$3.\pm 10.$&$1.\pm 6.$&$37.\pm 7.$&$16.\pm 2.$\\
&&&&&($2.\pm5.$)&&($13.\pm 4. $)&\\

2020-02-20&10:00:00-10:39:59&$4.2\pm0.1$&$406.\pm 3.$&$4.8\pm 0.1$&$4.\pm 1.0$&$1.3\pm 0.10$&$31.\pm 7.$&$11.4\pm 0.1$\\
&&&&&($0.55\pm0.06$)&&($5.4\pm0.3 $)&\\

2020-02-21&16:00:00-17:59:59&$6.0\pm0.8$&$476.\pm 8.$&$6.1\pm 0.4$&$4.0\pm 3.$&$1.3\pm 0.8$&$53.\pm 4.$&$17.\pm 2.$\\
&&&&&($1.2\pm0.9$)&&($15.\pm3. $)&\\

2020-02-23&17:40:00-19:09:59&$4.2\pm0.2$&$434.\pm 5.$&$3.5\pm 0.2$&$2.6\pm 0.4$&$1.0\pm 0.1$&$31.\pm 2.$&$12.1\pm 0.3$\\
&&&&&($0.24\pm0.05$)&&($3.2\pm0.5$)&\\

2020-02-27&08:10:00-08:59:59&$3.8\pm0.2$&$347.\pm 3.$&$5.8\pm 0.2$&$3.8\pm 0.5$&$1.5\pm 0.2$&$23.\pm 1.$&$9.1\pm 0.1$\\
&&&&&($0.4\pm0.1$)&&($1.9\pm0.3 $)&\\

2020-02-28&15:10:00-18:19:59&$6.\pm1.$&$390.\pm 20.$&$7.8\pm 0.8$&$3.\pm 6.$&$1.\pm 2.$&$34.\pm 2.$&$16.\pm 2.$\\
&&&&&($1.\pm1.$)&&($9.\pm 2. $)&\\

2020-03-01&20:50:00-21:29:59&$4.3\pm0.4$&$407.\pm 4.$&$3.1\pm 0.3$&$2.1\pm 0.4$&$0.8\pm 0.1$&$30.\pm 2.$&$11.9\pm 0.7$\\
&&&&&($0.4\pm0.2$)&&($5.\pm1. $)&\\

2020-03-06&14:50:00-15:29:59&$6.7\pm0.7$&$370.\pm 8.$&$7.6\pm 0.4$&$2.0\pm 0.6$&$1.0\pm 0.3$&$28.\pm 2.$&$14.0\pm 0.5$\\
&&&&&($0.7\pm0.3$)&&($7.\pm2. $)&\\

2020-03-07&15:00:00-19:29:59&$5.9\pm0.4$&$367.\pm 7.$&$4.7\pm 0.5$&$1.6\pm 0.3$&$0.7\pm 0.1$&$29.\pm 3.$&$11.8\pm 0.7$\\
&&&&&($0.17\pm0.08$)&&($2.2\pm0.4 $)&\\

2020-03-08&20:50:00-21:54:59&$3.7\pm0.4$&$337.\pm 2.$&$11.\pm 1.$&$6.\pm 2.$&$1.8\pm 0.6$&$17.\pm 1.$&$5.4\pm 0.3$\\
&&&&&($0.6\pm0.1$)&&($1.5\pm0.2 $)&\\

2020-03-20&15:20:00-16:19:59&$4.8\pm0.2$&$418.\pm 5.$&$4.6\pm 0.2$&$3.1\pm 0.4$&$1.2\pm 0.1$&$37.\pm 3.$&$14.4\pm 0.6$\\
&&&&&($0.3\pm0.1$)&&($4.0\pm0.6 $)&\\

2020-04-23&13:40:00-16:19:59&$4.1\pm0.1$&$394.\pm 4.$&$4.9\pm 0.4$&$2.8\pm 0.4$&$1.0\pm 0.1$&$24.\pm 2.$&$8.9\pm 0.5$\\
&&&&&($0.18\pm0.03$)&&($1.7\pm0.2 $)&\\
2020-04-27&02:00:00-02:59:59&$4.1\pm0.3$&$421.\pm 4.$&$3.9\pm 0.4$&$3.3\pm 0.9$&$1.0\pm 0.2$&$35.\pm 5.$&$11.3\pm 0.6$\\
&&&&&($0.5\pm0.2$)&&($5.6\pm0.8 $)&\\

2020-04-30&16:00:00-17:19:59&$3.9\pm0.1$&$291.\pm 2.$&$6.4\pm 0.2$&$2.7\pm 0.3$&$1.4\pm 0.1$&$16.\pm 1.$&$8.3\pm 0.3$\\
&&&&&($0.4\pm0.2$)&&($2.\pm1. $)&\\

2020-05-01&12:20:00-15:09:59&$4.4\pm1.0$&$303.\pm 4.$&$10.\pm 2.$&$5.\pm 40.$&$3.\pm 20.$&$18.\pm 2.$&$11.5\pm 0.6$\\
&&&&&($0.9\pm0.5$)&&($2.8\pm0.5 $)&\\

2020-05-04&03:50:00-05:09:59&$4.7\pm0.5$&$300.\pm 4.$&$8.4\pm 0.7$&$2.5\pm 0.5$&$1.2\pm 0.2$&$15.\pm 2.$&$7.6\pm 0.5$\\
&&&&&($0.34\pm0.06$)&&($2.1\pm0.3 $)&\\

2020-05-04&18:30:00-19:59:59&$3.1\pm0.5$&$317.\pm 3.$&$4.1\pm 0.3$&$5.\pm 2.$&$2.6\pm 1.0$&$28.\pm 2.$&$14.1\pm 0.4$\\
&&&&&($1.3\pm0.5$)&&($6.0\pm0.7 $)&\\
         
    \end{tabularx}
  \end{table}

\pagebreak

\end{document}